\documentclass[a4paper,fleqn]{cas-sc}

\usepackage[authoryear]{natbib}

\usepackage[english]{babel}
\usepackage[utf8]{inputenc}
\usepackage[hyphens, spaces, obeyspaces]{xurl}
\usepackage{amsmath,amsthm, amssymb, latexsym}
\usepackage{enumerate}
\usepackage{tikz}
\usetikzlibrary{shapes.geometric, arrows,backgrounds,fit}
\usepackage{booktabs}
\usepackage{bm}
\usepackage{csquotes}
\usepackage{float}
\usepackage[ruled,vlined]{algorithm2e}
\usepackage{graphicx}
\usepackage{subfig}

\usepackage{tikz}
\usetikzlibrary{arrows.meta, positioning}
\tikzset{%
  >={Latex[width=2mm,length=2mm]},
    minsize/.style={minimum width=15mm, minimum height=10mm},
    textwidth/.style={text width=15mm},
            base/.style = {rectangle, rounded corners, draw=black,
                           minimum width=3cm, minimum height=0.75cm,
                           text centered, font=\sffamily},
            base_box/.style = {rectangle, draw=gray,
                           minimum width=3cm, minimum height=0.75cm,
                           text centered, font=\sffamily},
  grid/.style = {base, fill=brown!70, text=black},
       obs/.style = {base_box, text=gray},
    kernel/.style = {base, fill=orange!70, text=black},
    prior/.style = {base, fill=blue!40, text=black},
         process/.style = {base, fill=gray!40, font=\ttfamily, text=black, minimum width=6.3cm}, 
    results/.style = {base, fill=orange!70, text=black},
}

\def\tsc#1{\csdef{#1}{\textsc{\lowercase{#1}}\xspace}}
\tsc{WGM}
\tsc{QE}
\tsc{EP}
\tsc{PMS}
\tsc{BEC}
\tsc{DE}

\begin{document}

\shorttitle{Direct sequential simulation for spherical linear inverse problems}
\shortauthors{M. Otzen et~al.}

\title[mode = title]{Direct Sequential Simulation for spherical linear inverse problems}

\tnotemark[1]

\tnotetext[1]{Github repository containing implementation is available at \url{github.com/mikkelotzen/spherical_direct_sequential_simulation}}

\author[1]{Mikkel Otzen}[type=editor,
auid=000,bioid=1,
prefix=, 
role=, 
orcid=0000-0001-6227-6916]
\cormark[1] 
\ead{mikotz@space.dtu.dk}

\author[1]{Christopher C. Finlay}
\author[2]{Thomas Mejer Hansen}

\address[1]{DTU Space, Centrifugevej 356, 2800 Kgs. Lyngby, Denmark}
\address[2]{Aarhus University, Høegh-Guldbergs Gade 2, building 1671, 8000 Aarhus C, Denmark}

\cortext[cor1]{Corresponding author}

\nonumnote{\textbf{Authorship statement.}
Mikkel Otzen: Conceptualization, Methodology, Software, Writing - original draft, review, and editing. Christopher C. Finlay: Supervision, Conceptualization, Methodology, Funding acquisition, Writing - review and editing. Thomas Mejer Hansen: Supervision, Conceptualization, Writing - review and editing.}

\begin{abstract}[S U M M A R Y]
We present a method for obtaining efficient probabilistic solutions to geostatistical and linear inverse problems in spherical geometry. Our Spherical Direct Sequential Simulation (SDSSIM) framework combines information from possibly noisy observations, that provide either point information on the model or are related to the model by a linear averaging kernel, and statistics derived from  a-priori training models.   It generates realizations from marginal posterior probability distributions of model parameters that are not limited to be Gaussian. We avoid the restriction to Cartesian geometry built into many existing geostatistical simulation codes, and work instead with grids in spherical geometry relevant to problems in Earth and Space sciences. 

We demonstrate our scheme using a synthetic example, showing that it produces realistic posterior realizations consistent with the known solution while fitting observations within their uncertainty and reproducing the model parameter distribution and  covariance statistics of a-priori training models. 
Secondly, we present an application to  real satellite observations, estimating the posterior probability distribution for the geomagnetic field at the core-mantle boundary.  Our results reproduce well-known features of the core-mantle boundary magnetic field, and also  allow probabilistic investigations of the magnetic field morphology.  Small-length scale features in the posterior realizations are not determined by the observations but match the covariance statistics extracted from geodynamo simulation training models. The framework presented here represents a step towards more general approaches to probabilistic inversion in spherical geometry.
\end{abstract}

\maketitle

\openup 1em 

\section{Introduction}
Globally-distributed Earth observation data is today available across many disciplines as low-Earth-orbit satellite missions, in combination with worldwide ground-based observing networks, provide a continuous stream of survey data. Such observations provide information and constraints on problems ranging from the impact of human activities on the Earth's surface and atmosphere  \cite[e.g.][]{jeong2017, jun2008} to inferring the structure and dynamics of the Earth's interior \cite[e.g.][]{Meschede_2015, Save_etal_2016, Gillet_2013}.  Common to such global problems is the need for analysis and interpretation on approximately spherical surfaces. Many analysis problems of this type can be formulated in terms of a linear inverse problem which connects observations, $\bm{d}$, to model parameters, $\bm{m}$, through a linear forward operator $\bm{G}$, i.e. $\bm{d} = \bm{G}\bm{m}$. We refer to solving such problems as linear inversion. Solution methods are traditionally based on least square methods \citep[e.g.][]{Menke_2018} and have in recent years been developed to include probabilistic solutions based on Bayesian methods \citep[e.g.][]{tarantola1982generalized, tarantola}. In the Bayesian formulation one seeks to estimate the posterior probability density function (pdf) of the model parameters, proportional to the product of an a priori pdf and a likelihood function. Probabilistic solutions to inverse problems with non-Gaussian prior information are often obtained using sampling methods such as the Metropolis algorithm, but this becomes very expensive when working with high dimensional model spaces. Here we present an alternative approach to generating realizations of the posterior pdf for linear inverse problems in spherical geometry, based on observations related to the model by a linear averaging kernel, that can account for non-Gaussian prior distributions for the model parameters. We also include the capability of using point data of the model parameters and refer to these as direct observations.

In geostatistics, solutions based on point data (where  the model parameters $m_i$ are known at some locations) are often obtained using kriging, i.e., interpolation through Gaussian process modelling conditioned on prior covariances, which provides the best linear unbiased predictions based on observed point data \citep{journel_mining, gslib}. Such kriging schemes can be extended to systems where there is a combination of point data and observations related to the model parameters by a linear averaging kernel, and to cases where only the latter are available \citep{thomas_linear_inv}. When the observation noise and a-priori pdfs are Gaussian, the posterior solution is provided by simple kriging through the mean and variance of a Gaussian estimate for each model parameter. Extensions of this framework, whereby the form of the posterior pdf can be obtained beyond means and covariances, is possible by sequentially simulating model parameters through sampling of local distributions conditional on prior information \citep{direct_sim}. Direct sequential simulation was applied to combinations of point and weighted linear average observations in Cartesian geometry in the VISIM algorithm of \cite{visim}.

In spherical geometry, to the best of our knowledge, no implementation of direct sequential simulation algorithms yet exists for obtaining probabilistic solutions of linear inverse problems, although the underlying methods are well understood. Recent work by \cite{alegr2020a} introduced a method for simulating Gaussian random fields on the d-dimensional unit sphere which is computationally very efficient but does not possess the non-Gaussian capabilities of direct sequential simulation. \cite{gneiting2013} analyses valid positive definite correlation functions on spheres which may be used to generate the necessary spherical covariance models. Spherical harmonics are a well known method to represent continuous fields on the sphere \citep{shtools}; they also provide a means to specify isotropic prior covariance functions on a sphere which have an exact correspondence to the spherical harmonic power spectra \citep[e.g.][]{Moritz_geodesy_1980, Jackson_1994, Hipkin_2001}.  Building on the work of \cite{visim} in Cartesian geometry, and making use of covariance models linked to spherical harmonic spectra, we implement direct sequential simulation in spherical geometry with the aim of providing a new tool for Earth and Space science problems. While we focus here on working with non-Gaussian posterior pdfs, sequential Gaussian simulation using point data is also possible with the tool presented. We illustrate our method by obtaining a probabilistic solution for the geomagnetic field at the Earth's core-mantle boundary, taking prior information from geodynamo simulations and based on real satellite observations. Our Spherical Direct Sequential Simulation (SDSSIM) algorithm enables probabilistic solutions to this problem without assuming a priori that the model parameters are Gaussian distributed.\\ 
In section \ref{sec:theory} we describe the linear forward problem, $\bm{d} = \bm{G} \bm{m}$, focusing on its discretization in spherical geometry. We next review the basic principles of Gaussian process based least-squares solutions to the inverse problem, sequential Gaussian simulation methods, and the theory of the direct sequential simulation method \citep{direct_sim, oz_dssim}. Section \ref{sec:implementation} gives a detailed description of the implementation of our SDSSIM algorithm. 
In section \ref{sec:case_geomag} we present the results of tests on both synthetic and real data, based on the geophysical problem of inferring the Earth's magnetic field at the core-mantle boundary from remote, noisy, satellite observations. Here we also demonstrate classic direct sequential simulation by using synthetic direct observations from a known simulation of the core-mantle boundary radial field. Finally we discuss the strengths and limitations of our method, along with our conclusions and some perspectives for future steps in \ref{sec:discussion}.

\section{Theory}
\label{sec:theory}

\subsection{The linear forward problem in spherical geometry}
\label{sec:theory_forward}

In spherical geometry, given observations, $d(\bm{r})$, and model parameters on a spherical surface, $m(\bm{s})$, related through a forward kernel operator $\mathcal{G}(\bm{r},\bm{s})$ we consider a linear forward problem of the form shown in equation (\ref{eq:dgm}).

\begin{align}
    d(\bm{r}) = \int_{S} \mathcal{G}(\bm{r},\bm{s}) m(\bm{s}) \enskip dS
\label{eq:dgm}
\end{align}

where $dS = \sin \theta' d\theta' d\phi'$, with $\bm{r} = (r,\theta,\phi)$ indicating locations of the observations, and $\bm{s} = (r',\theta',\phi')$ indicating locations of the model parameters on a spherical surface of radius $r'$. This system describes observations that are related to the model parameters by a linear averaging kernel. The integral equation in (\ref{eq:dgm}) may be approximated numerically via quadrature rules; here we use a Gauss-Legendre quadrature scheme appropriate for spherical geometry \citep[e.g.][]{atkins_glq, shtools}, in which the integration is carried out on a $(2N_{q}-1) \times N_{q}$ grid, where $N_{q}$ is the number of latitudinal nodes. $\cos{\theta'}$ are then the Gauss-Legendre nodes on the interval $[-1,1]$, with corresponding integration weights, $w_s$. $\phi'$ is chosen such that the points are equally spaced with separation $\pi/(N_q-1/2)$ on the interval $[0,2\pi[$. For model parameters on a sphere distributed according to Gauss-Legendre quadrature rules, this numerical integration is exact for polynomials of degrees less than $2N_q$ \citep{atkins_glq}. The integral may then be discretized according to equation (\ref{eq:glq_rule}).

\begin{align}
    d(\bm{r}) = \frac{\pi}{N_q-1/2}\sum_{i=1}^{N_m} w_{i} \mathcal{G}(\bm{r},\bm{s}_i) m(\bm{s}_i)
    \label{eq:glq_rule}
\end{align}

where $N_m = (2N_{q}-1) \times N_{q}$ is the number of model parameters. For a series of $N_d$ observations, $\bm{d} =\\ \big[d_1(\bm{r}_1),\hdots, d_i(\bm{r}_i), \hdots, d_{N_d}(\bm{r}_{N_d})\big]^T$, with a vector of model parameters, $\bm{m} = \big[m_1(\bm{s}_1),\hdots, m_i(\bm{s}_i), \hdots, m_{N_m}(\bm{s}_{N_m})\big]^T$; we absorb the constant, $\frac{\pi}{N_q-1/2}$, and integration weights $w_i$ into the elements of a matrix $\bm{G}$ (size $N_d \times N_m$) such that

\begin{align}
    G_{ij} = \frac{\pi}{N_q-1/2} w_i \mathcal{G}(r_j,s_i)
    \label{eq:glq_forward_param}
\end{align}

Other grids on the sphere could alternatively be used, along with suitable quadrature weights.  We adopted the Gauss-Legendre grid for simplicity and due to the ease of transforming to a spherical harmonic representation. Any linear forward problem in spherical geometry may then be written in the familiar form

\begin{align}
    \bm{d} = \bm{G} \bm{m}
    \label{eq:glq_rule_vec}
\end{align}

Here we are concerned with the inverse problem of how best to estimate $\bm{m}$, a vector of parameter values on a spherical surface grid, given noisy observed data $\bm{d}$ linearly related to the model, along with any prior information on the model parameters.

\subsection{Equivalent least-squares solution to the linear inverse problem}
\label{sec:theory_equiv_lsq}

A simple solution to the above inverse problem exists if we are able to assume the spherical surface model parameters can be represented by a Gaussian probability density function (pdf) with a priori mean $\bm{\mu}_0$ and covariance $\bm{C}_m$, while the observations, $\bm{d}$, represent realizations of Gaussian random variables with data error covariance $\bm{C}_e$ \citep{tarantola}. The least-squares solution is then also a Gaussian pdf with mean

\begin{align}
    \bm{\hat{m}}_{LSQ} = \bm{\mu}_0 + \bm{C}_m \bm{G}^T \bm{S}^{-1} \bigg(\bm{d} - \bm{G} \bm{\mu}_0\bigg)
\end{align}

and covariance

\begin{align}
    \bm{\hat{C}}_{LSQ} = \bm{C}_m - \bm{C}_m \bm{G}^T \bm{S}^{-1} \bm{G} \bm{C}_m
\end{align}

where

\begin{align}
    \bm{S} = \bm{C}_e + \bm{G} \bm{C}_m \bm{G}^T
\end{align}

This solution is identical to the solution of a simple kriging system \citep{thomas_linear_inv}. 
Before presenting spherical direct sequential simulation (section \ref{sec:theory_dsim}) as an alternative solution method which avoids these often restrictive Gaussian assumptions, we first briefly describe the method of sequential Gaussian simulation on the sphere.

\subsection{Sequential Gaussian Simulation on the sphere}
\label{sec:theory_gauss}
The method of sequential simulation \citep[e.g.][]{gslib, visim} involves inferring Gaussian posterior realizations $\bm{\hat{m}}$ of the random variables $\bm{m}$, from the observations $\bm{d}$.
For a joint distribution of $N_m$ random variables, $m_i$, conditioned on a set of known observations, $\bm{d}$, the $N_m$ variate cumulative distribution function (cdf) is 
\begin{align}
\begin{split}
F_{\bm{m}}(m_1,...,m_{N_m}|\bm{d}) &= P\{m_i \geq \hat{m}_i, i = 1,...,{N_m} | \bm{d} \}\\ 
&= P\{m_1 \geq \hat{m}_1 | \bm{d} \} P\{m_2 \geq \hat{m}_2 | \bm{d}, \hat{m}_1 \} \hdots P\{m_N \geq \hat{m}_N | \bm{d}, \hat{m}_1,\hat{m}_2, ...,  \hat{m}_{N-1}) \}
\end{split}
\label{eq:cdf_seqsim}
\end{align}
where $P$ denotes probability.  Sequential simulation involves drawing an $N_m$ variate sample based on ($\ref{eq:cdf_seqsim}$) making use of the product rule of probability, such that each probability term on the right-hand side is sampled in succession \citep{gslib}. Realizations are thus obtained in a series of $N_m$ sequential steps, gradually increasing the conditioning,
beginning with the observations, $\bm{d}$. 

For a Gaussian random field, drawing samples satisfying  (\ref{eq:cdf_seqsim}) equates to drawing from the Gaussian pdf, $\mathcal{N}(\mu_k,\sigma_k^2)$, where $\mu_k$ and $\sigma_k^2$ are the kriging mean and variance found by solving the kriging system \citep[e.g.][]{journel_mining, gslib,visim}, which in our notation is
\begin{align}
    \sum_{i = 1}^{N_v} C_v(v_i,v_j) \lambda_i = c_{vm}(v_i,\hat{m}_k) \quad \forall j = 1,\dots,N_v \qquad \mbox{or} \qquad  \bm{C}_v \bm{\lambda} = \bm{c}_{vm}
    \label{eq:covsys_init}
\end{align}

where $v_i$ is a member of the $N_v$ available conditional variables in each step (observations, point data, and previously simulated model parameters), $\hat{m}_k$ is the model parameter currently being simulated, and $\lambda_i$ are known as the kriging weights which determine the desired Gaussian pdf, $\mathcal{N}(\mu_k,\sigma_k^2)$. $C_v(v_i,v_j)$ are the a priori covariances between conditional variables including any measurement error covariance, and $c_{vm}(v_i,m_t)$ are the covariances between conditional variables and the target model parameter. Solving equation (\ref{eq:covsys_init}) for the kriging weights $\bm{\lambda}$, the kriging mean and variance are
\begin{align}
    \mu_k &= \bm{\lambda}\cdot(\bm{v} - \mu_0\bm{\bar{e}}) + \mu_0 \quad \text{where} \quad \bm{\bar{e}} = \big[1, \hdots, 1\big]^T \text{of length $N_v$} 
    \label{eq:kmean}\\
    \sigma^2_k &= \sigma^2_0 - \bm{\lambda} \cdot \bm{c}_{vm}
    \label{eq:kvar}
\end{align}
where $\mu_0$ and $\sigma_0^2$ are a-priori estimates of the mean and variance of the model parameters. 

Covariances between observation pairs and observation/model parameter pairs can be obtained from the forward relation between the observation and model parameters defined in (\ref{eq:glq_rule}) and (\ref{eq:glq_rule_vec}), the a-priori model parameter covariance measure, $C\left\{ \right\}$, 
and an example a-priori model, $m_0(\bm{s}_i)$. With the covariance of an observation pair defined by (\ref{eq:glq_rule}), and again absorbing the constant and integration weights into $G(\bm{r}_p,\bm{s}_i)$, we have
\begin{align}
\begin{split}
    C_{dd}\big\{d(\bm{r}_p),d(\bm{r}_q)\big\} &= C\Bigg\{\sum_{i=1}^{N_m} G(\bm{r}_p,\bm{s}_i) m_0(\bm{s}_i), \sum_{j=1}^{N_m} G(\bm{r}_q,\bm{s}_j) m_0(\bm{s}_j)\Bigg\}\\ 
    &= \sum_{i=1}^{N_m} \sum_{j=1}^{N_m} G(\bm{r}_p,\bm{s}_i) G(\bm{r}_q,\bm{s}_j) C\big\{m_0(\bm{s}_i), m_0(\bm{s}_j)\big\}
\end{split}
\label{eq:obs_cov}
\end{align}
and
\begin{align}
    C_{dm}\big\{d(\bm{r}_p),m_0(\bm{s}_q)\big\} = \sum_{i=1}^{N_m} G(\bm{r}_p,\bm{s}_i) C\big\{m_0(\bm{s}_i), m_0(\bm{s}_q)\big\}
\label{eq:obsmod_cov}
\end{align}

All required covariances are thus available given an a priori model parameter covariance based on prior information regarding the random variable on the spherical surface and knowledge of the forward problem.
The kriging system (\ref{eq:covsys_init}) can therefore be expanded as follows, in order to explicitly show the contributing parts of the covariance matrix
 
\begin{align}
    \begin{bmatrix}
        \bm{C}_{dd} + \bm{C}_e  & \bm{C}_{dm}\\
        \bm{C}_{dm}^T & \bm{C}_{m}\\
    \end{bmatrix}
\bm{\lambda} = 
    \begin{bmatrix}
    \bm{c}_{dm}\\
    \bm{c}_{mm}
    \end{bmatrix}    
    \label{eq:DSSIM_krigsys}
\end{align}

$\bm{C}_{dd}$ is a matrix of observation to observation covariances computed using (\ref{eq:obs_cov}), $\bm{C}_e$ holds observation data error covariances. $\bm{C}_{m}$ and the vector $\bm{c}_{mm}$ contain covariances between previously simulated model parameters and between previously simulated model parameters and the target model parameter, both are obtained directly from the a-priori model covariance. $\bm{C}_{dm}$ and $\bm{c}_{dm}$ are covariances from observations to previously simulated model parameters, and to the target model parameter respectively, as given by (\ref{eq:obsmod_cov}).\\
Solving the kriging system sequentially using the above covariances, results in a sequential Gaussian simulation model realization  $\mathcal{N}(\mu_k,\sigma_k^2)$. Example model realizations are drawn from the posterior distribution by visiting model parameters in a random order for each realization. 

Such sequential Gaussian simulation schemes are well-known and widely used in geostatistics.  However many physical processes in Earth and Space physics are fundamentally nonlinear which results in non-Gaussian statistics for the model parameters $\bm{m}$.  In the next section we extend the above treatment to permit non-Gaussian model parameter distributions, based on the method of direct sequential simulation \citep{journel_uncertainty,tran_dss,oz_dssim}.  This approach ensures the linear relationship (\ref{eq:glq_rule_vec}) between $\bm{d}$ and $\bm{m}$ is preserved, which is not the case for sequential Gaussian simulations after transforming  $\bm{d}$ and/or $\bm{m}$ to Gaussian variables. In sequential Gaussian simulation such a transformation destroys the linear relationship so the necessary covariance matrices cannot be expressed simply as a function of $\bm{C}_{mm}$  and $\bm{G}$.

\subsection{Direct Sequential Simulation for non-Gaussian fields on a sphere}
\label{sec:theory_dsim}

The system described in section \ref{sec:theory_gauss} allows one to sequentially simulate model parameters on a spherical surface given observations, leading to a realization of a Gaussian random field which fits the observations to within measurement error, and as far as this fit allows, reproduces the mean and covariance of the a priori information. We now go further and simulate non-Gaussian random fields using direct sequential simulation with histogram reproduction for a given training model, following the methods proposed by  \cite{journel_uncertainty}, \cite{tran_dss}, and \cite{oz_dssim}.\\
A normal-score transform of a training model to variables, $\bm{y}$, that follow a standard Gaussian distribution (i.e. zero mean, variance one), and the associated back-transformation to original values may be performed through (\ref{eq:norm_score_trans}) and (\ref{eq:norm_score_trans_back}).

\begin{align}
    \bm{y} = H^{-1}(F(\bm{m_0)})
    \label{eq:norm_score_trans}\\
    \bm{m_0} = F^{-1}(H(\bm{y}))
    \label{eq:norm_score_trans_back}
\end{align}

Where $H^{-1}$ is a standard Gaussian quantile function with cdf, $H$, $F$ is the training model cdf with quantile function $F^{-1}$, and $\bm{m}_0$ are the training model values. This transformation describes the connection between random variables with a Gaussian distribution, and non-Gaussian distributions defined by the training model. It allows one to generate a collection of non-Gaussian cdf's through which the sampling in the sequential simulation steps described in equation (\ref{eq:cdf_seqsim}) can occur. Generating a non-Gaussian cdf is achieved by substituting the standard Gaussian representation of the training model, $\bm{y}$, for a Gaussian distribution, $\bm{y}_n$, with mean, $\mu_n$, and variance $\sigma^2_n$. This can be achieved as follows

\begin{align}
    \bm{y}_n = H^{-1}(\bm{u})\sigma_n + \mu_n
    \label{eq:loccond_lutgen}
\end{align}

through inverse transform sampling using a vector, $\bm{u}$, of $N_u$ uniformly spaced quantiles between zero and one, which divides the Gaussian distribution into intervals of equal probability. The ranges of the mean and variance should cover approximately $[-3.5,3.5]$ and $[0,2]$ respectively, to fully utilize the training model in characterizing conditional distributions \citep{oz_dssim}. A variance range of [0,2] is used in order to alleviate cases where the obtained kriging variances lie outside the domain of the generated local conditional distributions as shown by \cite{deutsch2001preliminary}.
Performing transformation with $\bm{y}_n$ as in (\ref{eq:norm_score_trans_back}) results in a discrete vectorized quantile function, $\bm{q}_n$, conditional on the training model and with length $N_u$ describing a distribution with mean, $\mu_i$, and variance, $\sigma^2_i$,

\begin{align}
    \bm{q}_n = F^{-1}(H(\bm{y}_n))
    \label{eq:loccond_vec}
\end{align}

from which a sample, $z_{i}$, can be drawn using a uniform distribution, $U$, discretized in $N_u$ intervals

\begin{align}
    z_{i} = \bm{q}_n\big(U(0,N_u)_i\big)
    \label{eq:loccond_vec_samp}
\end{align}

Solving the kriging system yields an estimated mean and variance of the local Gaussian distributions and a distribution is then assigned from $\bm{q}_n$ based on the mean, $\mu_i$, and variance, $\sigma^2_i$, closest to the kriging mean and variance. In this step a distance measure must be used and our implementation is described in section \ref{sec:implement_conddist}.
We refer to the chosen distributions collectively as the local distributions. However, reproduction of the training model is only ensured if the applied local distribution has mean and variance equal to the kriging mean and variance \citep{journel_uncertainty}. This further requires that the local distributions are scaled to have exactly the kriging mean and variance. For a value sampled from one of the local distributions, this is achieved by

\begin{align}
    \hat{m}_k &= (z_{i} - \mu_{i}) \cdot \frac{\sigma_k}{\sigma_{i}} + \mu_k
    \label{eq:scaleZ}
\end{align}

where $\hat{m}_k$ is the final simulated model parameter value that makes up the vector of model parameters $\bm{\hat{m}}_{DSS} = \big[\hat{m}_1, \hdots, \hat{m}_k, \hdots, \hat{m}_{N_m} \big]^T$ in a given realization.
A probabilistic solution is achieved by collecting $N_p$ model parameter realizations in the matrix, $\bm{\hat{M}}_{DSS}$, and computing the sample covariance as follows

\begin{align}
    \bm{\hat{C}}_{DSS} = \frac{1}{N_p - 1} \big(\bm{\hat{M}}_{DSS} - \bm{\hat{\mu}}_{DSS}\bm{\bar{e}}\big) \big(\bm{\hat{M}}_{DSS} - \bm{\hat{\mu}}_{DSS}\bm{\bar{e}}\big)^T \quad \text{where} \quad \bm{\bar{e}} = \big[1, \hdots, 1\big] \text{of length $N_p$} 
    \label{eq:cov_dss}
\end{align}

where $\bm{\hat{\mu}}_{DSS}$ is an $N_m$ length column vector of the model parameter means. This procedure ensures that simulations represent samples from the posterior probability density function of the model parameters based on the mean, variance, covariance structure, and histogram provided by the training model, while honoring the data \citep{tran_dss, oz_dssim}.

\section{Implementation}
\label{sec:implementation}
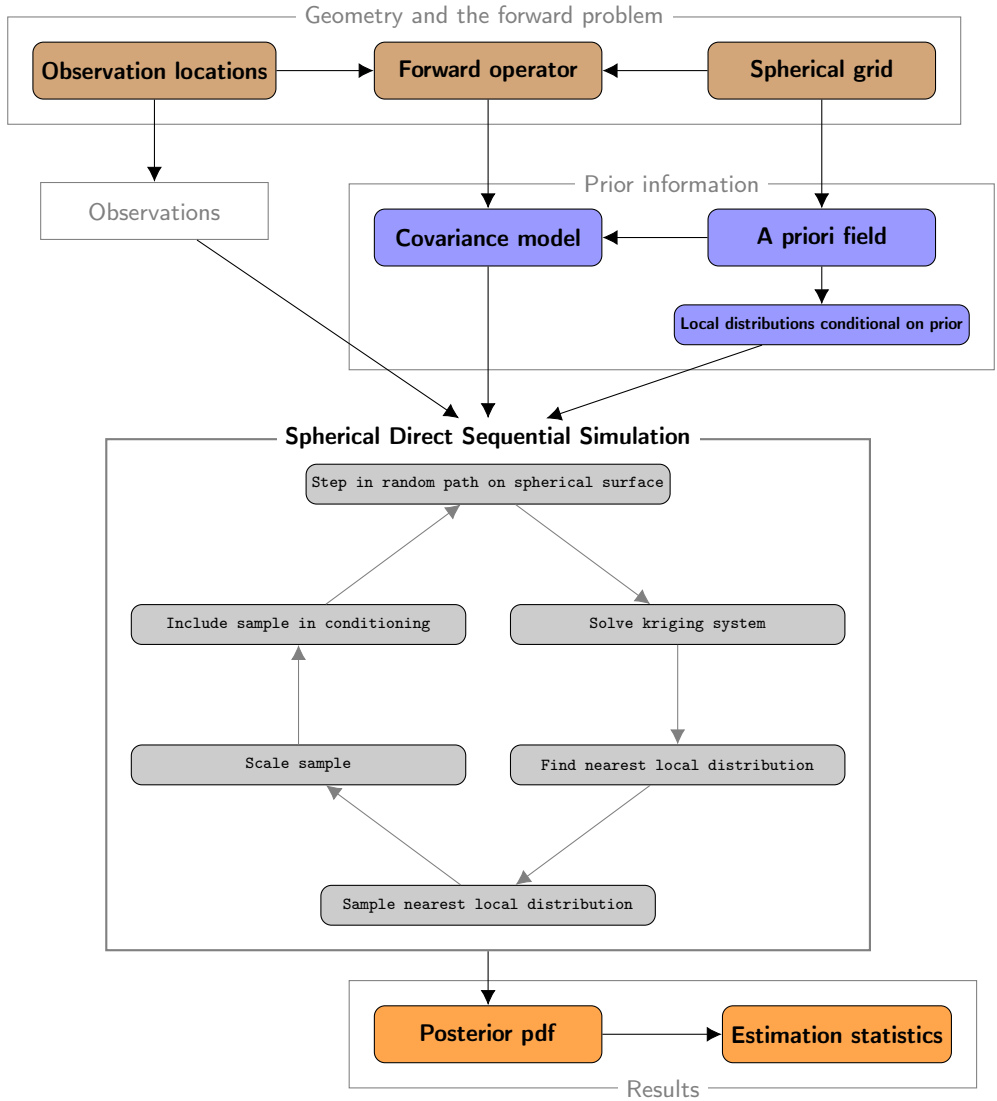
\begin{figure}[!ht]
\centering
\begin{tikzpicture}[node distance=2.2cm, every node/.style={}, font=\sffamily, align=center]
  \node (grid) [grid, xshift=4.4cm] {\textbf{Spherical grid}};
  \node (prior) [prior, below of = grid] {\textbf{A priori field}};
  \node (condtab) [prior, below of = prior, scale = 0.7, xshift=0.0cm, yshift=1.5cm] {\textbf{Local distributions conditional on prior}};
  \node (obs_grid) [grid, xshift=-4.4cm] {\textbf{Observation locations}};
  \node (kernel) [grid, left of = grid, xshift=-2.2cm] {\textbf{Forward operator}};
  \node (obs) [obs, below of = obs_grid, yshift=0.35cm] {Observations};
  \node (cov) [prior, below of = kernel] {\textbf{Covariance model}};
  
  \node (rand) [process, below of = cov, scale = 0.7, yshift = -1.5cm] {\textbf{Step in random path on spherical surface}};
  \node (kriging) [process, below of = rand, xshift=2.5cm, scale = 0.7, yshift = 0.5cm] {\textbf{Solve kriging system}};
  \node (conddist) [process, below of = kriging, scale = 0.7, yshift = 0.5cm] {\textbf{Find nearest local distribution}};
  \node (condlut) [process, below of = conddist, xshift=-2.5cm, scale = 0.7, yshift = 0.5cm] {\textbf{Sample nearest local distribution}};
  \node (addcond) [process, below of = rand, xshift=-2.5cm, scale = 0.7, yshift = 0.5cm] {\textbf{Include sample in conditioning}};
  \node (ozscale) [process, below of = addcond, scale = 0.7, yshift = 0.5cm] {\textbf{Scale sample}};
  
  \node (model) [results, below of = condlut, yshift=0.5cm] {\textbf{Posterior pdf}};
  \node (eval) [results, right of = model, xshift=2.4cm] {\textbf{Estimation statistics}};
  
  \node (box) [draw, thick, inner xsep=1em, inner ysep=1em, fit= (rand) (kriging) (conddist) (condlut) (addcond) (ozscale)] {};
  \node (sdssim) [fill = white, text=black] at (box.north) {\textbf{Spherical Direct Sequential Simulation}};
  
  \node (box_prior) [draw, inner xsep=1em, inner ysep=1em, fit= (prior) (condtab) (cov)] {}; 
  \node (box_prior_text) [fill = white, text=gray] at (box_prior.north) {Prior information};
  
  \node (box_grid) [draw, inner xsep=1em, inner ysep=1em, fit= (grid) (obs_grid)] {};
  \node (box_grid_text) [fill = white, text=gray] at (box_grid.north) {Geometry and the forward problem};

  \node (box_results) [draw, inner xsep=1em, inner ysep=1em, fit= (model) (eval)] {};
  \node (box_results_text) [fill = white, text=gray] at (box_results.south) {Results};

  \path[->] (cov) edge[color = black] (sdssim);
  \path[->] (box) edge[color = black] (model);
  \path[->] (model) edge[color = black] (eval);
  \path[->] (condtab) edge[color = black] (sdssim);
  \path[->] (prior) edge[color = black] (condtab);
  \path[->] (grid) edge[color = black] (prior);
  \path[->] (grid) edge[color = black] (kernel);
  \path[->] (prior) edge[color = black] (cov);
  \path[->] (obs) edge[color = black] (sdssim);
  \path[->] (obs_grid) edge[color = black] (kernel);
  \path[->] (obs_grid) edge[color = black] (obs);
  \path[->] (kernel) edge[color = black] (cov);
  \path[->] (rand) edge (kriging); 
  \path[->] (kriging) edge (conddist);
  \path[->] (conddist) edge (condlut); 
  \path[->] (condlut) edge (ozscale); 
  \path[->] (ozscale) edge (addcond);
  \path[->] (addcond) edge (rand); 
\end{tikzpicture}
  \caption{The implementation of Spherical Direct Sequential Simulation (SDSSIM) shown as a flowchart. We consider our implementation as five distinct modules. Geometry and forward operator to set up the spherical inverse problem, prior information as required to solve the inverse problem, observations conditioning the solution, SDSSIM performing inversion by solving system equations, and results as the final output.}
  \label{char:SDSSIM_flowchart}
\end{figure}

We have implemented the methods described in section \ref{sec:theory} as a Python repository called Spherical Direct Sequential Simulation, which is available on Github at \url{github.com/mikkelotzen/spherical\_direct\_sequential\_simulation}. The implementation includes five modules. (i) The geometry of the problem and the forward operator, relating the observations to the model parameters, (ii) the prior information, (iii) the measured observations, (iv) the simulation itself, and (v) the posterior pdf output. Figure \ref{char:SDSSIM_flowchart} gives an overview of these modules; their content is described in more detail below. The propagation of information is shown with arrows.

\begin{figure}[ht!]
	\centering
		\includegraphics[scale=.4, angle=-90]{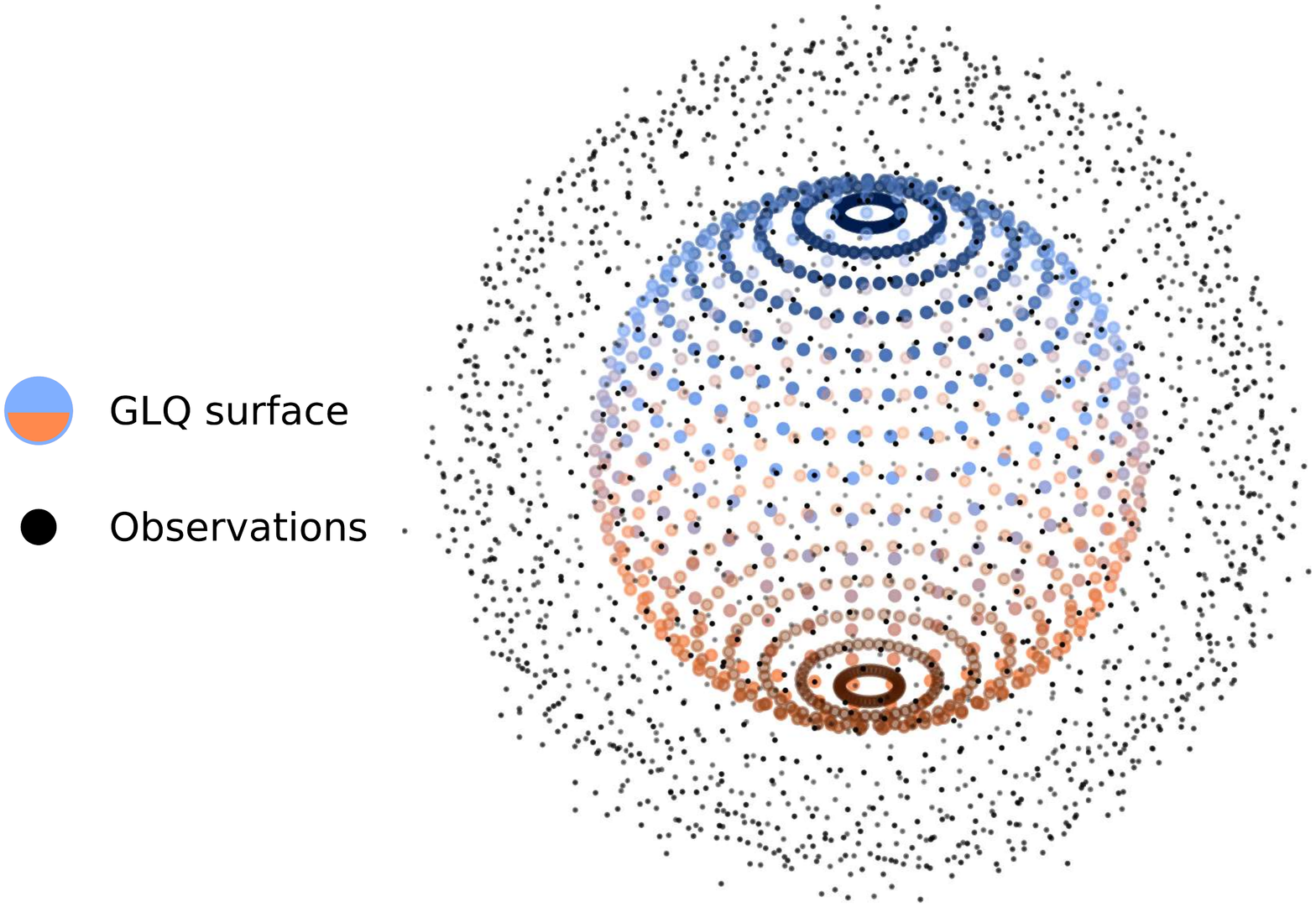}
	\caption{Illustration of unit Gauss-Legendre quadrature grid and observations distributed according to an equal area grid with varying radii in spherical space. It is also possible to work with other grids, including approximate equal area grids, depending on the application. Grids on spherical surfaces are typical of Earth observation and geophysics problems utilizing satellites or global ground networks for data collection.}
	\label{fig:implementation_geometry}
\end{figure}

\subsection{Geometry and the forward problem}
In SDSSIM the spherical polar coordinates for the surface points to be modelled and observations are stored in arrays of length $N_m$ and $N_d$ respectively, with the radius of the points to be modelled on the spherical surface of interest being a constant, $r'$.
Equation (\ref{eq:dgm}) defines the spherical linear inverse problem by connecting random variables on a spherical surface to observations. This spatial structure is illustrated with an example in figure \ref{fig:implementation_geometry}, where a unit radius Gauss-Legendre quadrature (GLQ) grid is displayed among equal area distributed observations with varying radii. The discretized forward operator, $\bm{G}$, which defines the connection between the observations and spherical surface in equations (\ref{eq:glq_forward_param}) - (\ref{eq:glq_rule_vec}), is an array of size $(N_d, N_m)$, and its construction is problem dependent. In section \ref{sec:case_geomag} an example of a forward operator is presented.

\subsection{Observations}
We store observations, $d(\bm{r})$, in the vector array $\bm{d}$ of length $N_d$, with one value for each observation coordinate $\bm{r} = (r,\theta,\phi)$. These values are noisy and linearly related to the desired model parameters being simulated as described by the forward problem in (\ref{eq:dgm})-(\ref{eq:glq_rule_vec}). In case of point data, i.e. direct observations of the model parameters themselves, these are simply added to the model parameter vector during simulation with the associated error added to the corresponding covariance indices. 
In section \ref{sec:case_geomag} we consider a case using the magnitude of the radial component of Earth's magnetic field at satellite altitude and a case using synthetic direct observations at the core-mantle boundary.

\subsection{Prior information}
Prior information is contained in the training model, $\bm{m}_0$, an array of length $N_m$. The values in the training model provides a distribution, with a priori mean, $\mu_0$, and variance, $\sigma^2_0$. The training model is also used as conditioning for a range of possible local distributions and further prior information in the form of a covariance model. Included in the implementation is the possibility of semi-variogram modelling. In our implementation this allows for estimating a covariance model from the training model based on the assumption that it is second-order stationary and isotropic. It is also possible to use an a priori power spectrum derived from the model to specify the covariance model.

\subsubsection{Covariance model}

For an isotropic field one can compute the spherical harmonic power spectrum and use this to define the covariance model, and hence the covariance matrix \citep[e.g.][in the geomagnetic framework]{Jackson_1994, Hipkin_2001}.
Given any such covariance matrix, $\bm{C}_m$, the data to data and data to model parameter covariances are computed through (\ref{eq:obs_cov}) and (\ref{eq:obsmod_cov}) respectively.
The observation to observation data error covariance, $\bm{C}_e$, will later be added to $\bm{C}_{dd}$; this is a diagonal array of the data error covariance level.

\subsubsection{Local distributions conditional on prior}
\label{sec:implement_conddist}
Local distributions conditional on the a priori training model used to sample model parameters are implemented as a lookup-table (LUT) based on equations (\ref{eq:norm_score_trans})-(\ref{eq:loccond_vec}). The procedure is shown as pseudo-code in algorithm \ref{alg:lut_Q}. The first input is the training model and the second is the number of local distribution quantiles, $N_u$. The number of quantiles is chosen by the user and should at most equal the size of the training model, as it controls the level of detail expressed in the conditional distributions, which cannot exceed the level of detail in the histogram on which they are based. Two further inputs are the discretization levels, $N_\mu$ and $N_\sigma$, the ranges of the mean and standard deviation used in generating Gaussian distributions as shown in equation (\ref{eq:loccond_lutgen}). The discretization level of the mean and standard deviation range determines the number of local distributions in the LUT.
From these inputs, a uniformly spaced array, $\bm{u}$, in the range zero to one is generated containing $N_u$ equally spaced values. $\bm{u}$ and the range of mean and standard deviation values determined by $N_\mu$ and $N_\sigma$ are then used iteratively in equation (\ref{eq:loccond_lutgen}) and (\ref{eq:loccond_vec}) to generate the local distribution LUT, $\bm{Q}$, of size $(N_u, N_\mu, N_\sigma)$. 
In our implementation the Python package scikit-learn \citep{scikit-learn} is used to handle the normal-score and inverse transformations, $F^{-1}()$ and $H^{-1}()$.\\
Having generated $\bm{Q}$, we require a measure for finding the nearest local distribution given a kriging mean and variance, $\mu_k$ and $\sigma^2_k$, such that a simulated model parameter (\ref{eq:scaleZ}) can be computed. This is achieved using an array $\bm{\Psi}$ of measures

\begin{align}
\bm{\Psi} = |\bm{Q_\mu}-\mu_k\bm{\bar{E}}|/\Delta m_0+|\bm{Q}_{\sigma^2}-\sigma^2_k\bm{\bar{E}}|/\sigma^2_0
\label{eq:distance_to_Q}
\end{align}

where $\Delta m_0$ is $max(\bm{m}_0)-min(\bm{m}_0)$, and $\sigma^2_0$ is the training model variance. $\bm{Q_\mu}$ and $\bm{Q}_{\sigma^2}$ are arrays of the mean and variance for each local distribution, their sizes are $(N_\mu, N_\sigma)$, and $\bm{\bar{E}}$ is an array of ones matching their size. $\bm{\Psi}$ is then likewise of size $(N_\mu, N_\sigma)$ and the index of the minimum value indicates the required nearest local distribution. Note that $N_\mu$ and $N_\sigma$ determines the size of the above computation. Setting these to very large values can lead to heavy computational cost, slowing the simulation down, as the above is carried out for each model parameter simulation.

\vspace{1em}
\begin{algorithm}[H]
\DontPrintSemicolon
\SetStartEndCondition{ }{}{}%
\SetKwProg{Fn}{def}{\string:}{}\SetKwFunction{Range}{range}
\SetKw{KwTo}{in}\SetKwFor{For}{for}{\string:}{}%
\SetKwIF{If}{ElseIf}{Else}{if}{:}{elif}{else:}{}%
\SetKwInOut{Input}{input}\SetKwInOut{Output}{output}
\SetAlgoLined
\Input{$\bm{m}_0$, $N_u$, $N_\mu$, $N_\sigma$}
\Output{$\bm{Q}$}
Generate $\bm{u}$ based on $N_u$\;
Generate $N_\mu$ mean values evenly spaced between $-3.5\dots3.5$\;
Generate $N_\sigma$ standard deviation values evenly spaced between $0.0\dots2.0$\;
From $\bm{m}_0$ compute $F^{-1}$\;
$i = 0$, $j = 0$\;
\For{$\mu$ in the range $-3.5\dots3.5$}{
\For{$\sigma$ in the range $0.0\dots2.0$}{
$\bm{y} = H^{-1}(\bm{u})\sigma + \mu$ \tcp*{Equation (\ref{eq:loccond_lutgen})}
$\bm{q} = F^{-1}(H(\bm{y}))$ \tcp*{Equation (\ref{eq:loccond_vec})}
$\bm{Q}_{ij} = \bm{q}.copy()$\;
$j+=1$}$i+=1$}
\caption{Generating a look-up table for the local distribution of the model parameters, conditional on the data and already simulated model parameters.}
\label{alg:lut_Q}
\end{algorithm}
\vspace{1em}

\subsection{Spherical Direct Sequential Simulation}
Expanding on the outline of SDSSIM given in the algorithm flowchart of figure \ref{char:SDSSIM_flowchart}, the algorithm contains the following steps.

\begin{enumerate}
    \setlength\itemsep{0.01em}
    \item Determine a random path through the model parameters on the spherical surface.
    \item At each location in the random path solve equation (\ref{eq:covsys_init}) with the appropriate covariance matrices based on all available observations and previously simulated values. The kriging mean, $\mu_k$, and variance, $\sigma_k^2$, are then determined through (\ref{eq:kmean}) and (\ref{eq:kvar}).
    \item The nearest local distribution in $\bm{Q}$ is found through (\ref{eq:distance_to_Q}). This provides the local distribution closest to the kriging mean and variance.
    \item Draw a sample from this nearest local distribution.
    \item Scale the sample through (\ref{eq:scaleZ}) such that it originates from a local distribution with mean and variance exactly equal to the kriging mean and variance.
    \item Add the scaled sample to the list of previously simulated model values for use in the rest of the simulation.
    \item 2.-6. is repeated until all model parameters have been visited.
\end{enumerate}

Performing the above with different random paths each time yields an ensemble of realizations from the posterior pdf, which are collected in the matrix $\bm{\hat{M}}_{DSS}$, and can then be used to estimate statistics such as the sample mean and covariance.
Algorithm \ref{alg:sdssim} shows pseudo-code for the SDSSIM algorithm in the case of computing $N_p$ realizations with the outputs $\bm{\hat{M}}_{DSS}$, $\bm{\hat{C}}_{DSS}$ and $\bm{\hat{\mu}}_{DSS}$. While not shown in algorithm \ref{alg:sdssim}, the implementation includes an option of skipping step 3-5 in the above, which results in spherical sequential Gaussian simulation.

\begin{algorithm}
\DontPrintSemicolon
\SetStartEndCondition{ }{}{}%
\SetKwProg{Fn}{def}{\string:}{}\SetKwFunction{Range}{range}
\SetKw{KwTo}{in}\SetKwFor{For}{for}{\string:}{}%
\SetKwIF{If}{ElseIf}{Else}{if}{:}{elif}{else:}{}%
\SetKwInOut{Input}{input}\SetKwInOut{Output}{output}
\SetAlgoLined
\Input{$N_p$, $\bm{d}$, $\mu_0$, $\sigma^2_0$, $\bm{Q}$, $\bm{Q}_{\sigma^2}$, $\bm{Q}_\mu$, $\Delta m_0$, $\bm{G}$, $\bm{C}_m$, $\bm{C}_{dd}$, $\bm{C}_{dm}$, $\bm{C}_e$}
\Output{$\bm{\hat{M}}_{DSS}$, $\bm{\hat{C}}_{DSS}$, $\bm{\hat{\mu}}_{DSS}$}
\;
Generate $\bm{\hat{M}}_{DSS}$ as empty array of size $(N_m,N_p)$\;
\For{$realization$ in \textbf{range}($0,N_p$)}{
\quad Set model parameter $path$ as random indices from $0$ to $N_m-1$\;
\quad Set $steps$ as empty list\;
\For{$step$ in $path$}{
\# Conditional variables for current step\;
\quad $\bm{v} = \text{stack}(\bm{d}, \bm{\hat{M}}_{DSS}[steps, realization])$\;
\# Compute RHS in (\ref{eq:covsys_init})\;
\quad $\bm{c}_{mm} = \bm{C}_{m}[step,steps]$\;
\quad $\bm{c}_{dm} = \bm{C}_{dm}[:,step]$\;
\quad $\bm{c}_{vm} = \text{stack}(\bm{c}_{dm},\bm{c}_{mm})$\;
\# Compute covariance part of LHS in (\ref{eq:covsys_init})\;
\quad $\bm{C}_{mm} = \bm{C}_{m}[steps,:][:,steps]$\;
\quad $\bm{C}_{dmm} = \bm{C}_{dm}[:,steps]$\;

\quad $ \bm{C}_{v} = \begin{bmatrix}
        \bm{C}_{dd} + \bm{C}_e  & \bm{C}_{dmm}\\
        \bm{C}_{dmm}^T & \bm{C}_{mm}\\
    \end{bmatrix}$

\# Solve kriging system for $\lambda$\;
\quad $\bm{\lambda} = \bm{C}_v^{-1}\bm{c}_{vm}$ \tcp*{Solved equation (\ref{eq:covsys_init})}
\# Compute kriging mean and variance\;
\quad $\mu_k = \bm{\lambda}\cdot(\bm{v} - \mu_0\bm{\bar{e}}) + \mu_0 \quad \text{where} \quad \bm{\bar{e}} = \big[1, \hdots, 1\big]^T \text{of length $N_v$}$ \tcp*{Equation (\ref{eq:kmean})}
\quad $\sigma^2_k = \sigma^2_0 - \bm{\lambda} \cdot \bm{c}_{vm}$ \tcp*{Equation (\ref{eq:kvar})}
\# Look-up the nearest local distribution\;
\quad $\bm{\Psi} = \text{abs}(\bm{Q}_\mu-\mu_k)/\Delta m_0+\text{abs}(\bm{Q}_{\sigma^2}-\sigma^2_k)/\sigma^2_0$\;
\quad $\text{nearest} = \text{argmin}(\bm{\Psi})$\;
\# Draw sample and scale to distribution with kriging mean and variance\;
\quad $z_{step} = \bm{Q}[U(0,N_u),\text{nearest}]$\;
\quad $\mu_{step} = \bm{Q}_\mu[:,\text{nearest}]$\;
\quad $\sigma_{step} = \text{sqrt}(\bm{Q}_{\sigma^2}[:,\text{nearest}])$\;
\quad $\hat{m}_{step} = (z_{step} - \mu_{step}) \cdot \frac{\sigma_k}{\sigma_{step}} + \mu_k$ \tcp*{Equation (\ref{eq:scaleZ})}
\# Update model parameter array and steps\;
\quad $\bm{\hat{M}}_{DSS}[step, realization] = \hat{m}_{step}$\;
\quad $steps.\text{append}(step)$}
}
\;
\# Compute sample mean and covariance\;
$\bm{\hat{\mu}}_{DSS} = \text{mean}(\bm{\hat{M}}_{DSS},axis=-1)$\;
$\bm{\hat{C}}_{DSS} = \frac{1}{N_p - 1} \big(\bm{\hat{M}}_{DSS} - \bm{\hat{\mu}}_{DSS}\bm{\bar{e}}\big) \big(\bm{\hat{M}}_{DSS} - \bm{\hat{\mu}}_{DSS}\bm{\bar{e}}\big)^T \quad \text{where} \quad \bm{\bar{e}} = \big[1, \hdots, 1\big] \text{of length $N_p$}$
\caption{Spherical Direct Sequential Simulation}
\label{alg:sdssim}
\end{algorithm}
\vspace{1em}

\clearpage
\section{A case study from geophysics: Core-mantle boundary magnetic field estimation}
\label{sec:case_geomag}
We now demonstrate SDSSIM on an example spherical linear inverse problem, estimating the radial component of Earth's magnetic field on the approximately spherical core-mantle boundary (CMB) from globally-distributed satellite magnetic observations \citep[e.g.][]{Langel_87, Bloxham_etal_89, Gubbins_2004, Finlay_2020}.

First we test our method on a synthetic case considering two distinct scenarios where different types of observations are generated from a known source, referred to below as the synthetic truth. In subsection \ref{subsec:synthsat} we consider synthetic satellite observations and in subsection \ref{subsec:directobs} we consider direct observations of some of the model parameters (i.e. of the synthetic truth radial field at the CMB with added noise).  In all cases, for prior information we use as traning models an ensemble of $487$ instances of the core-mantle boundary field (up to spherical harmonic degree 30) from a numerical model of the magnetic field generating dynamo process in Earth's outer core \citep{aubert_gastine_fournier_2017}. These dynamo fields contain highly localized field structures that lead to a more Laplacian than Gaussian distribution of the radial field at the CMB. The synthetic truth is chosen to be another snapshot from the dynamo model, not included in the training set.

Having validated the method in this test case we go on to use real satellite magnetic field observations from Swarm Alpha, one satellite from ESA's \textit{Swarm} constellation mission \citep[e.g.][]{swarm_mission}. \textit{Swarm} data are freely available and were downloaded through the Swarm Virtual Research Environment\footnote[1]{Swarm Virtual Research Environment  \url{swarm-vre.readthedocs.io}}.\\
The geomagnetic forward problem is of the form (\ref{eq:dgm}) with a forward operator based on the Green's function describing a potential field solution to Laplace's equation in spherical geometry for internal source Neumann boundary conditions \citep[e.g.][]{Gubbins_Roberts_83, hammer_kernel}. For simplicity we consider only observations of the radial component of the field;  the forward operator linking the radial geomagnetic field at an observation location, $\bm{r} = (r, \theta,\phi)$, to the radial field at a source location on the spherical core-mantle boundary , $\bm{s} = (r', \theta', \phi')$, is then

\begin{align}
\mathcal{G} = \frac{1}{4 \pi} \frac{h^2 (1-h^2)}{f^3}
\end{align}

where

\begin{align*}
h =\frac{r'}{r}, \quad f = \frac{\sqrt{r^2+{r'}^2-2r r'\cos\Upsilon}}{r} \quad \text{with} \quad \cos\Upsilon = \cos\theta \cos\theta' + \sin\theta \sin\theta'\cos(\phi-\phi')
\end{align*}

We represent the radial magnetic field in physical space at the CMB at a radius of $r'=3480.0\text{km}$, on a Gauss Legendre quadrature grid with $N_q = 31$ latitudinal nodes and thus have $N_m = 1891$ model parameters. This allows accurate transformation to a spherical harmonic representation up to degree $n=30$.

When only satellite observations are available we use the training ensemble of dynamo model realizations to specify an a-priori covariance function for the CMB radial magnetic field model.  Assuming isotropy and stationarity over the spherical surface the covariance function for the radial magnetic field may be written \citep{Jackson_1994, Hipkin_2001}

\begin{align}
C_{B_r}(\Upsilon) = \sum_{n=1}^\infty \frac{n+1}{2n+1} R_n(r') P_n(\cos \Upsilon)
\label{eq:cov_spec}
\end{align}

where $R_n(r')=(n+1)  \left( \frac{a}{r'}\right) ^{2n+4} \sum\limits_{m=0}^{n} (g^m_n)^2 + (h^m_n)^2$ is the Lowes spherical harmonic power spectrum \citep{Lowes_1966} at $r'$, with  $g^m_n$ and $h^m_n$ Schmidt quasi-normalized spherical harmonic coefficients of degree $n$ and order $m$, $a$ is the reference radius of the spherical harmonics, and $P_n$ are Legendre polynomials of degree $n$. We used (\ref{eq:cov_spec}) to compute an a-priori model covariance matrix between all grid points on the CMB that is consistent with the isotropic second order statistics given by the power spectra of the realizations in the dynamo model training ensemble.  The mean of these defines our a-priori model covariance matrix, $C_m$. The training models in the dynamo ensemble are provided to spherical harmonic degree $n=30$, whereas equation (\ref{eq:cov_spec}) involves a summation to infinity. Generating a covariance matrix using (\ref{eq:cov_spec}) based on truncation to degree 30 can thus result in non-positive-definite covariance matrices due to the truncation. To avoid this problem we implement a function that gradually tapers the spectra to zero beyond degree $30$, using $f_{taper} = 0.5e^{-5n}+0.5e^{-2n}$. The a-priori local distribution for the model parameters is obtained by concatenating histograms from the training ensemble of numerical dynamo simulations of the CMB radial field. Figure \ref{fig:core_synth_rep}(b) shows the training ensemble histograms and (c) shows the ensemble of Lowes power spectra used to generate the a priori covariance model. 

For the test based on direct observations (i.e. on sampled values of the radial magnetic field at the CMB with noise added), the a-priori covariance is instead specified using an exponential semi-variogram model estimated from the available direct observations and a-priori local distributions are generated from their distribution shape.

\subsection{Probabilistic inversion of synthetic test data}

\subsubsection{Synthetic satellite observations}
\label{subsec:synthsat}
For this test we used $N_p = 2773$ synthetic observations located at satellite altitude at positions sampled along real Swarm Alpha orbits taken from April-June 2018. To the synthetic radial field computed on the satellite orbits (based on the synthetic truth model) we add zero mean random Gaussian noise with std.dev. of $2\text{nT}$. In the simulation we characterize this with a diagonal data error covariance matrix of $(2\text{nT})^2$.

\begin{figure}[ht!]
	\centering
		\includegraphics[scale=.6]{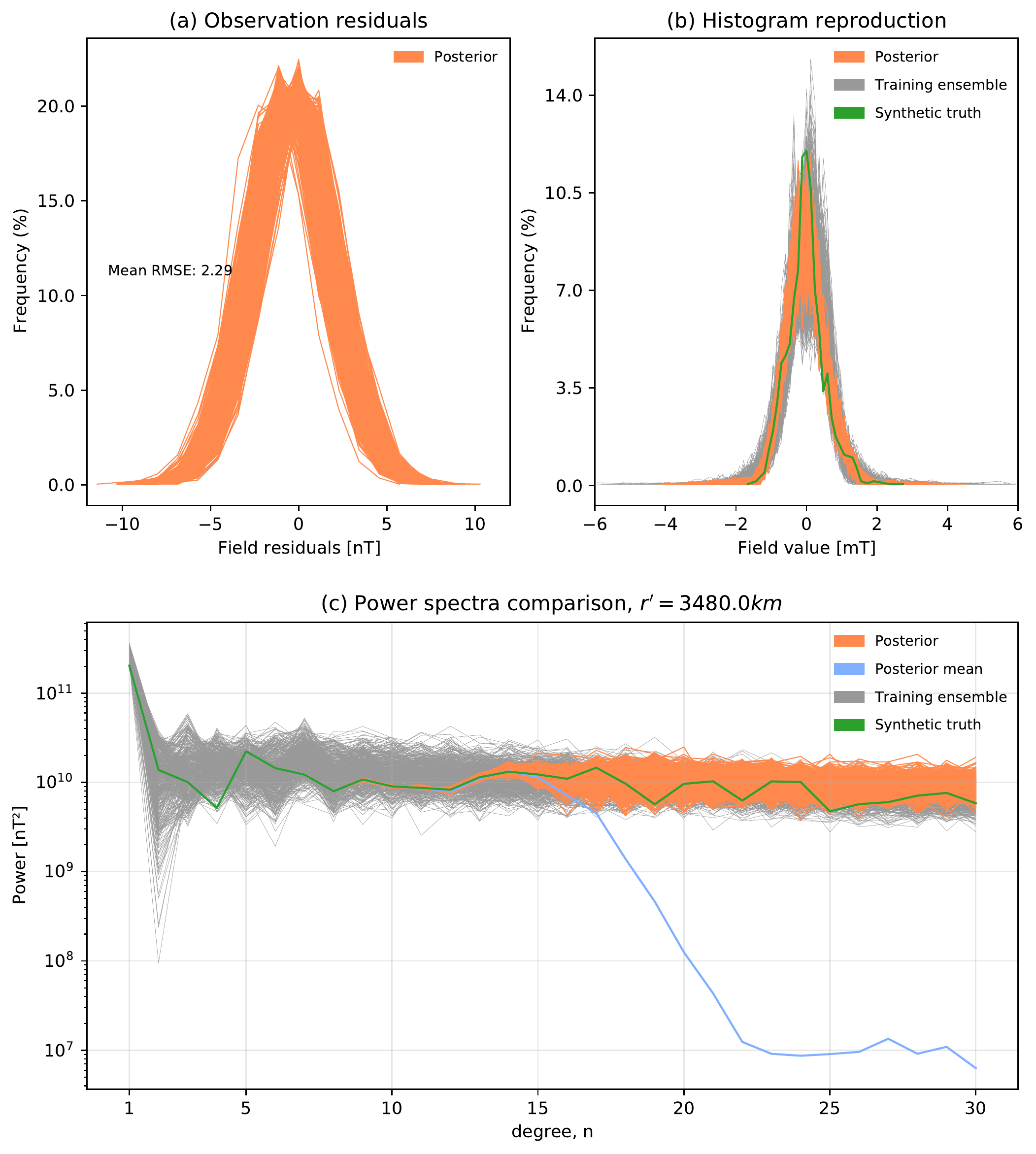}
	\caption{Overview of diagnostics of core field estimation for the synthetic test case with observations related to the model parameters by the linear averaging kernel, $\mathcal{G}$. Results are shown for 1000 realizations conditional on synthetic observations generated from a known synthetic truth core field. (a) Posterior realization fits to the synthetic observations. (b) and (c) are respectively histograms and the spherical harmonic power spectra (which specifies the model covariance function via (\ref{eq:cov_spec})), of posterior realizations (orange) compared with the training ensemble (grey) and synthetic truth (green).}
	\label{fig:core_synth_rep}
\end{figure}

\begin{figure}[ht!]
	\centering
		\includegraphics[scale=.6]{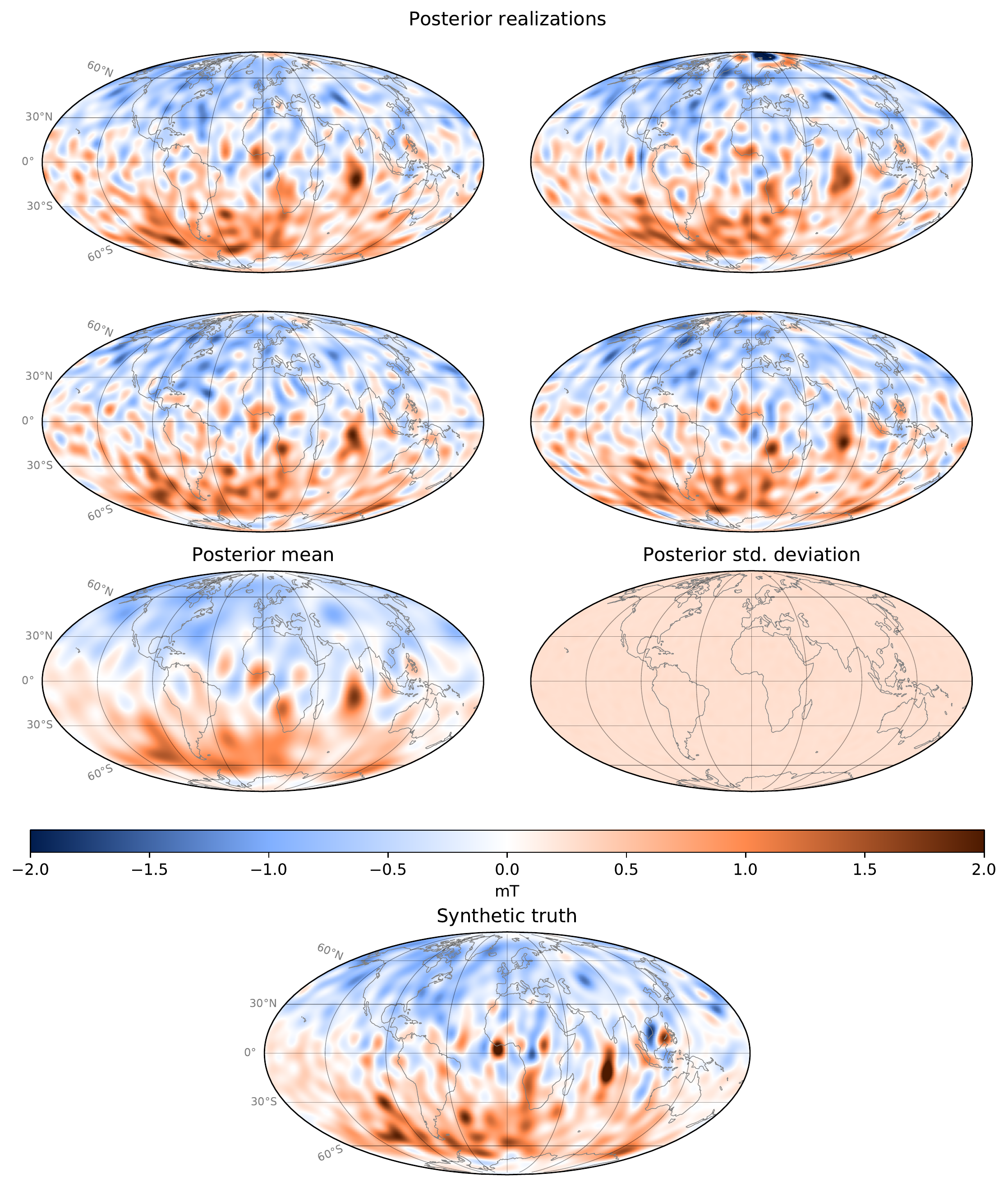}
	\caption{Posterior realization samples for core estimation using synthetic satellite observations as well as the posterior mean and std. deviation compared to the synthetic truth used in this test.}
	\label{fig:core_synth_tile}
\end{figure}

Figure \ref{fig:core_synth_rep} summarizes simulation diagnostics after 1000 posterior realizations are generated. The residual histograms in (a) demonstrate the posterior realizations fit the observations to a level similar to the added noise. In (b) the histogram of synthetic truth model values is well reproduced by the posterior realizations, and the posterior is within the training ensemble. Some posterior realizations exhibit larger magnitudes than the synthetic truth as seen by the distribution tails. The power spectra in (c) displays close agreement between the posterior and synthetic truth until around degree 8. At higher degrees the posterior is more broadly distributed around the synthetic truth, reaching similar width as the training ensemble by degree 17. Already at degree 15 the spectrum of the posterior mean starts to drop, indicating a point at which the small scales due to the prior start to average out across realizations.
Sample maps of the posterior realizations as well as the posterior mean, standard deviation, and the synthetic truth are shown on figure \ref{fig:core_synth_tile}. As expected we see differences in the small scale features between posterior realizations. In comparison the synthetic truth is smoother, except the high amplitude features which are more concentrated. These results make it clear that with the prior covariance model, number of synthetic observations, and Gaussian noise used in this test, we are able to retrieve the synthetic truth only up to spherical harmonic degree 15. We further observe that the posterior standard deviation obtained with globally distributed synthetic satellite observations is uniform.

\begin{figure}[ht!]
	\centering
		\includegraphics[scale=0.53]{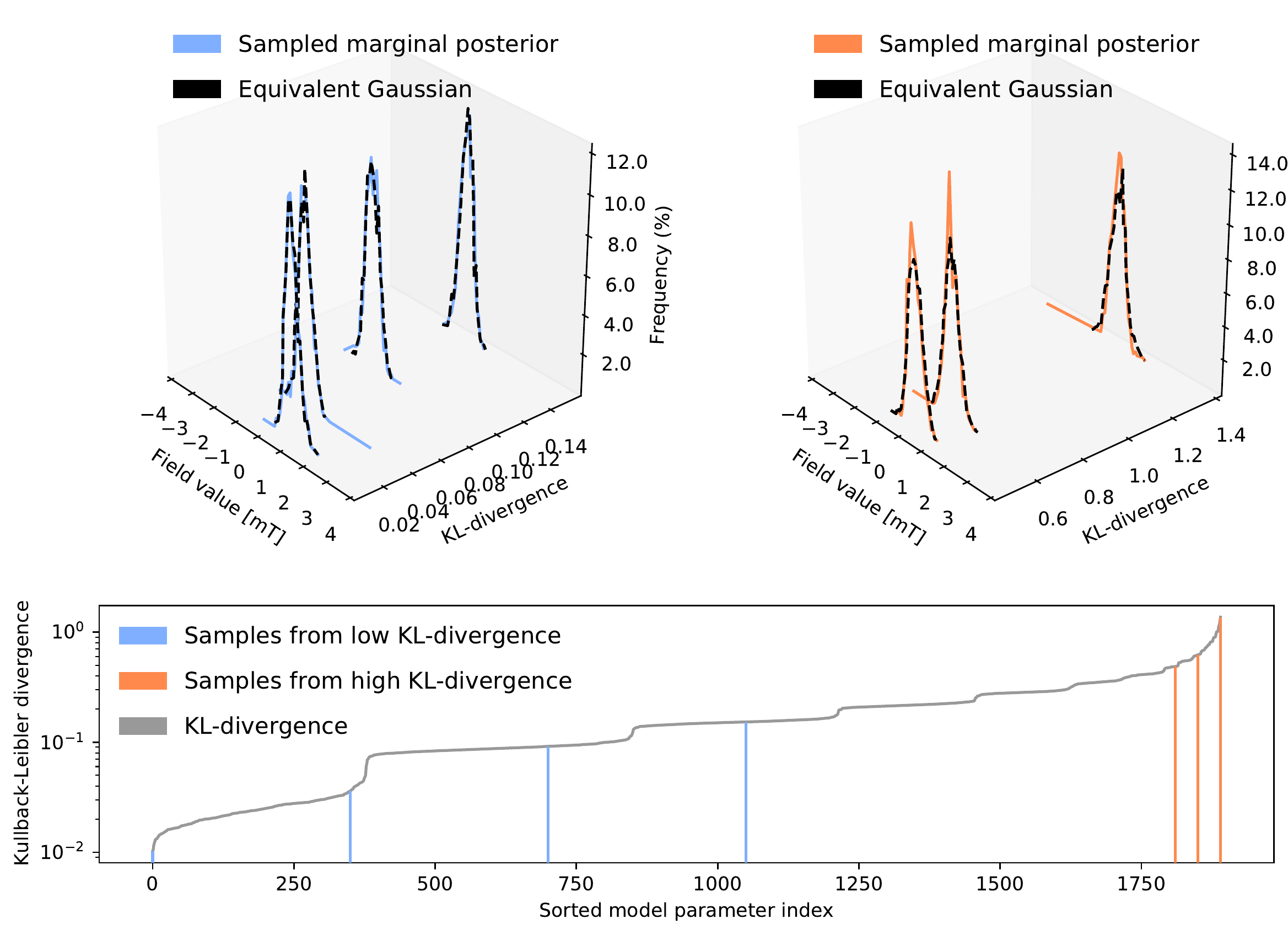}
	\caption{Examples of marginal posterior distributions based on low and high values of KL-divergence with respect to Gaussian distributions of equal mean and variance. Low values indicate high similarity. The blue distributions are selected examples with close to their equivalent Gaussian and the orange distributions selected examples diverging from their equivalent Gaussian. The grey curve shows the KL-divergence of all points on the CMB grid. The grid points have been sorted according to increasing values of KL-divergence.}
	\label{fig:core_synth_local}
\end{figure}

A characteristic output of SDSSIM is the local marginal posterior distribution for each model parameter in physical space on the spherical surface. We refer to these as the marginal posterior distributions. They each contain all the generated values from the posterior realizations at a specific location. Examples are presented in figure \ref{fig:core_synth_local}, selected based on their departure from a Gaussian distribution, as measured by the  Kullback-Leibler (KL) divergence \citep{kullback1951},
\begin{align}
    D_{KL} = \sum_{\hat{m}_k} P(\hat{m}_k) \log\bigg(\frac{P(\hat{m}_k)}{Q(\hat{m}_k)}\bigg)
\end{align}
where P is here the marginal posterior distribution and Q is a Gaussian distribution with equal mean and variance sampled the same number of times.  This allows us to identify marginal posterior distributions most similar and dissimilar to Gaussian distributions. Example marginal posterior distributions with low and high KL-divergence are selected in respectively blue and orange, along with their equivalent Gaussian distributions in black. Here it is clear that the marginal posteriors contain both distributions close to Gaussian and distributions with much sharper peaks and longer tails. SDSSIM is thus clearly capable of generating non-Gaussian probabilistic model parameter estimates.

\subsubsection{Synthetic direct observations of the CMB magnetic field}
\label{subsec:directobs}
\begin{figure}[ht!]
	\centering
		\includegraphics[scale=.6]{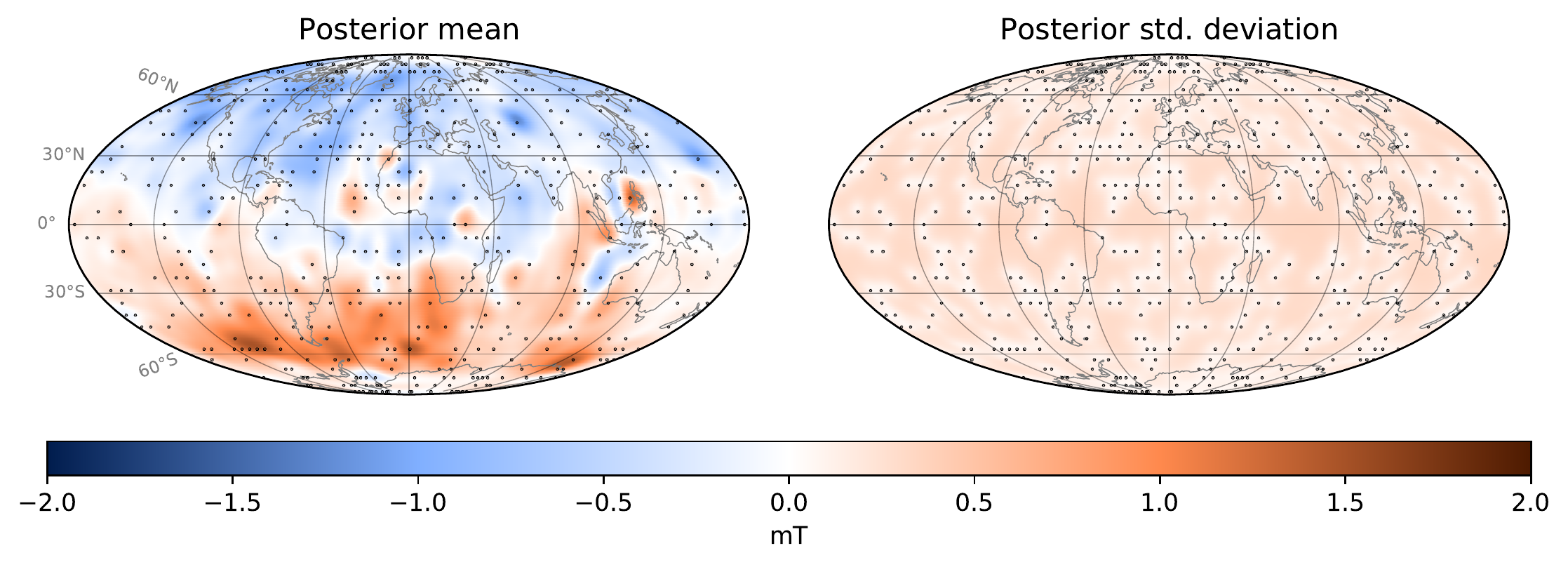}
	\caption{Posterior mean and std. deviation for synthetic core estimation using direct observations. The sampled direct observation locations are included as black dots.}
	\label{fig:core_synth_tile_direct}
\end{figure}

Validation tests based on synthetic direct observations of the model parameters are common in geostatistical studies. We report here briefly the results of such a test in order to demonstrate that our method can also be used within a more conventional geostatistical setup .  To obtain direct observations, we sampled $27\%$ of the synthetic truth radial magnetic field at the CMB and  added zero mean random Gaussian noise with a std.dev. of $2\text{nT}$.  1000 posterior realizations are generated using the covariance models and a-priori histograms described above. 
In figure \ref{fig:core_synth_tile_direct} we display the resulting posterior mean and std. deviation maps of the radial magnetic field at the CMB. The large scale structures are well reproduced for areas containing direct observations, whereas areas without direct observations lack the structures present in the synthetic truth (bottom of figure \ref{fig:core_synth_tile}). The posterior std. deviation clearly aligns with the locations of the sampled direct observations and show that these areas are more informed.
The histogram and semi-variogram reproduction is successful (not shown). This test shows that SDSSIM is capable of performing classical direct sequential simulation using only direct observations of the model.

\subsection{Probabilistic inversion of real satellite magnetic observations}
\label{sec:case_geomag_obs}
We now move on to the more interesting case of inferring the radial magnetic field at the CMB using real satellite data.  We use data from \textit{Swarm} Alpha sampled at 5 minute intervals, during dark times, over one year from 01-11-2018 to 01-11-2019, applying standard data selection criteria to remove observations collected during periods of high solar-driven field disturbances \citep{Kauristie_2017, Finlay_2020}.

\begin{figure}[ht!]
	\centering
		\includegraphics[scale=.6]{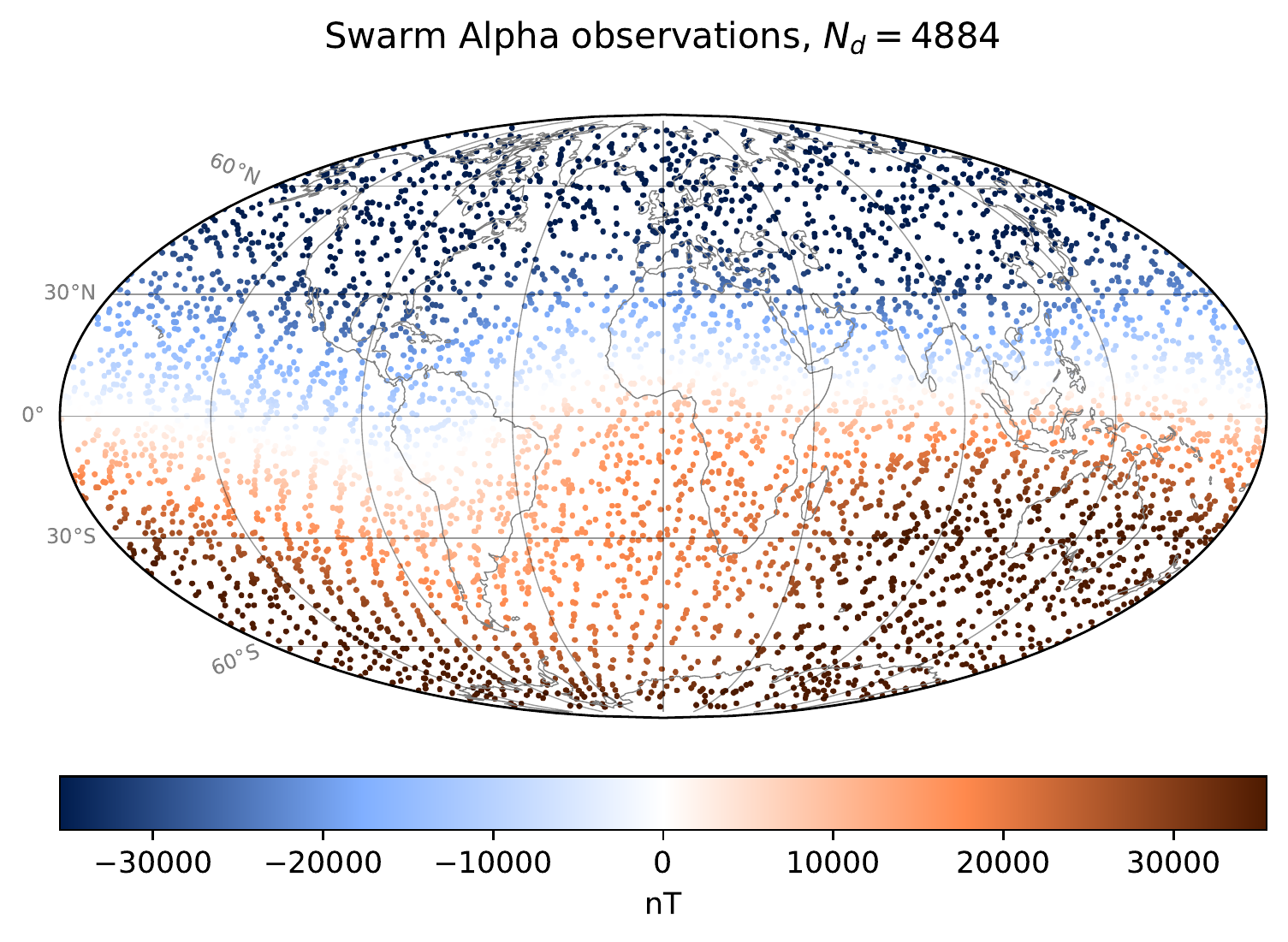}
	\caption{A year of Swarm Alpha observations of the radial magnetic field sampled at 5 minute intervals from 01-11-2018 to 01-11-2019. Observations have been selected based on dark and quiet sun-driven disturbance conditions. Models were used to remove contributions from the lithosphere, magnetosphere, and core field secular variation.}
	\label{fig:core_real_obs}
\end{figure}

This resulted in $N_d = 4884$ observations at altitudes between $432\text{km}$ and $452\text{km}$. In order to isolate the core field,  we removed the LCS-1 \citep{LCS1} model of the lithospheric magnetic field and the magnetospheric field and secular variation (changes over time of the core field) predicted by the CHAOS-7.2 model \citep{CHAOS7}. The resulting radial field observations are presented in figure \ref{fig:core_real_obs}.  For simplicity we assume the data error to be independent, Gaussian, with zero mean, and a standard deviation of $6\text{nT}$ at all latitudes.

\begin{figure}[ht!]
	\centering
		\includegraphics[scale=.6]{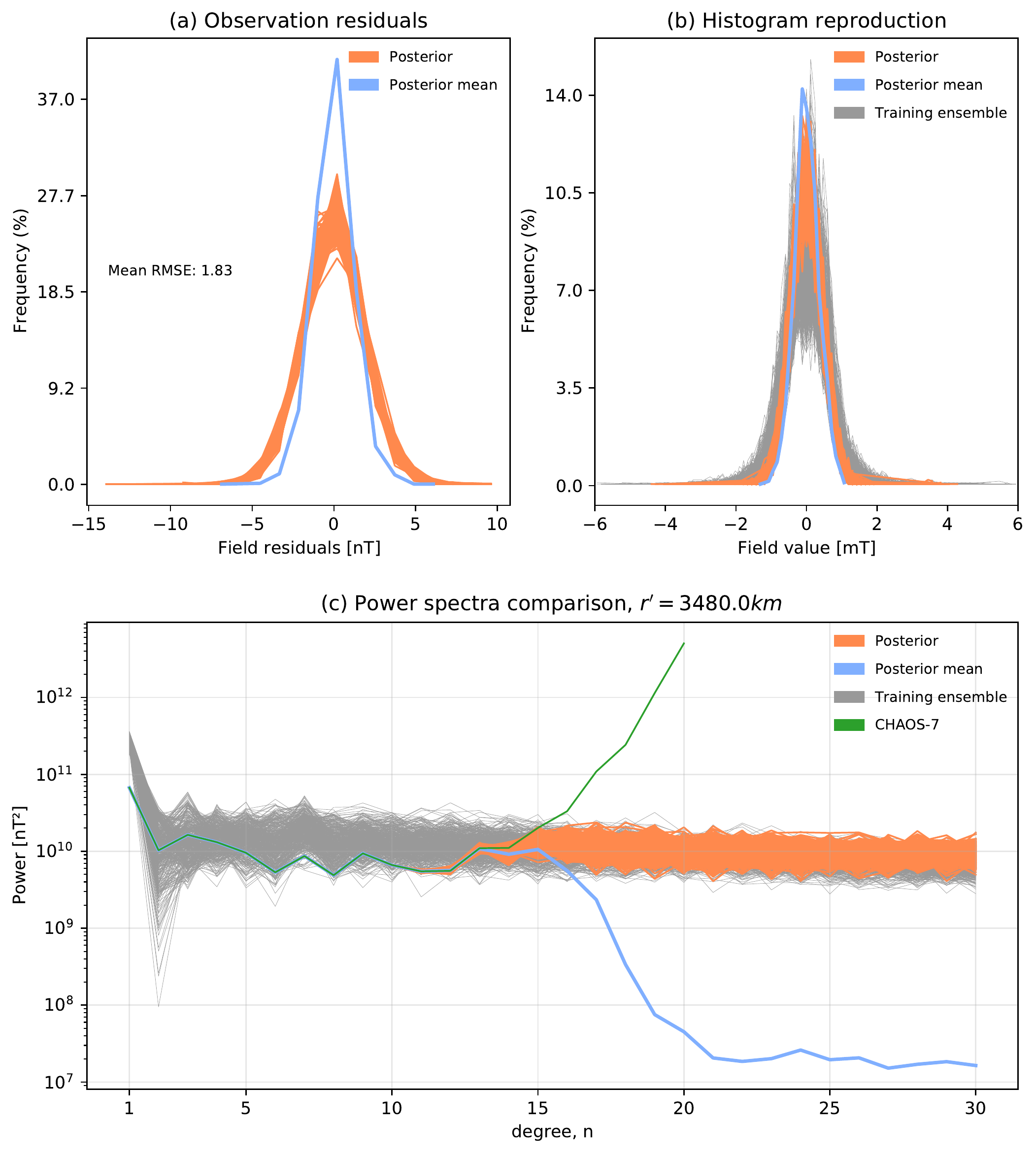}
	\caption{Overview of diagnostics of core field estimation from real satellite magnetic data. Results are shown for 500 realizations conditional on observations from the Swarm Alpha satellite. (a) Posterior realization fits to the observations. (b) The posterior realization histogram along with the training ensemble (c) Spherical harmonic power spectra at the core-mantle boundary of posterior realizations and their mean compared to the CHAOS-7 geomagnetic field model.}
	\label{fig:core_real_rep}
\end{figure}

Simulation diagnostics based on 500 posterior realizations are presented in figure \ref{fig:core_real_rep}. In (a) we see a fit of the posterior realizations to the observations with a mean RMSE of $1.83\text{nT}$, and also the fit of the posterior mean model, which was not explicitly constructed to fit the data. These are well below the assumed data error level of $6\text{nT}$ suggesting this was over-estimated. (b) shows the histograms of the radial field at the core-mantle boundary from posterior realizations compared to the posterior mean and training ensemble. The posterior distribution is narrower than the training ensemble with smaller tail amplitudes.  In (c) we compare power spectra of the posterior realizations and posterior mean to those of the training ensemble and the internal (lithosphere and core) part of the CHAOS-7 model for 01-11-2018. The posterior realizations agree very closely with CHAOS-7 until degree 10 at which point the spread in the ensemble of posterior realizations begins to broaden. The posterior mean matches CHAOS-7 until degree 13, after which it diverges as it contains the lithospheric field.  The posterior mean retains power until degree 15, beyond this it loses power since smaller scales in the posterior realizations are largely based on the prior covariance function which average out.

\begin{figure}[ht!]
	\centering
		\includegraphics[scale=.6]{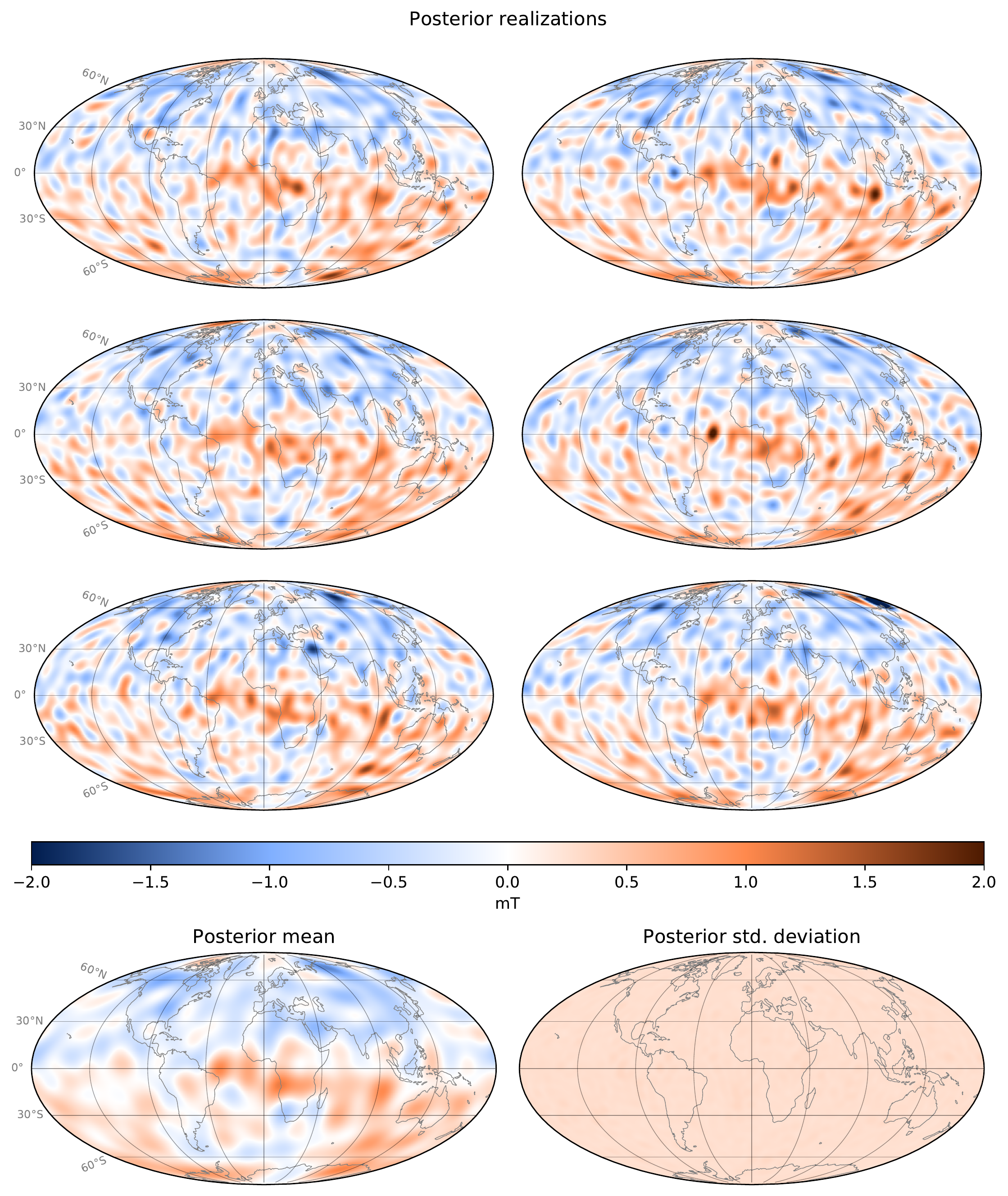}
	\caption{Posterior realization samples, mean, and std. deviation for core estimation based on $N_d=4884$ real satellite observations.}
	\label{fig:core_real_tile}
\end{figure}

\begin{figure}[ht!]
	\centering
	\includegraphics[scale=.8]{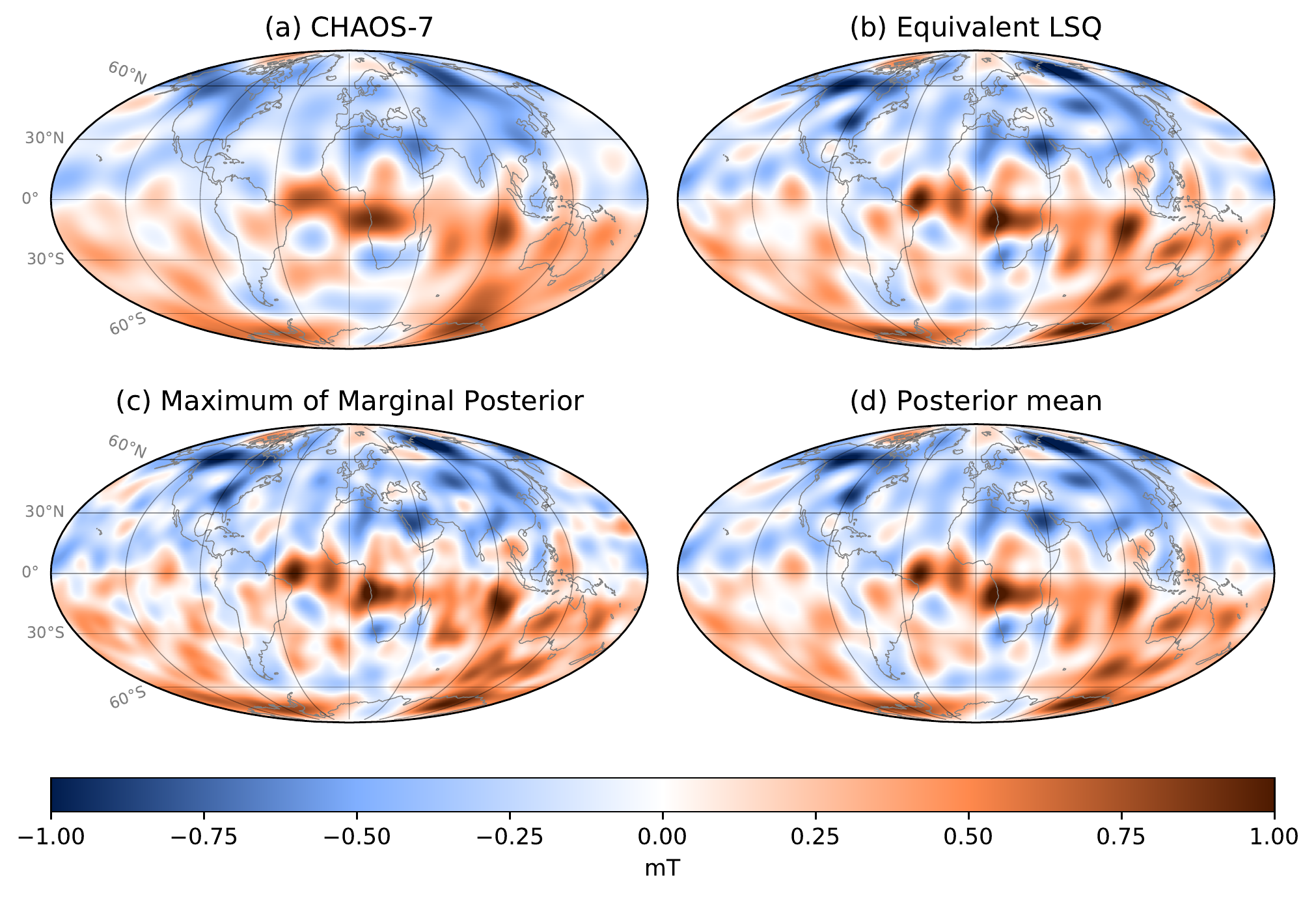}
    \caption{Field map comparison of the CHAOS-7 radial geomagnetic core to the posterior mean, the maximum of the marginal posterior, as well as the equivalent least-squares solution. CHAOS-7 is shown up to degree 13 while the rest is shown to degree 30.}
    \label{fig:mean_vs_chaos}
\end{figure}

Example posterior realizations, as well as the posterior mean and standard deviation are shown on figure \ref{fig:core_real_tile}. The posterior mean is smooth in comparison to the realizations since it has little power beyond degree 15.  The observations used in this experiment clearly do not constrain the posterior beyond degree 15.  Figure \ref{fig:mean_vs_chaos} compares maps of the CMB radial field from CHAOS-7, maps of the equivalent least-squares solution and  mean of the posterior realizations, and a map collecting the radial field values from the maximum of the marginal posterior distribution at each grid point.
CHAOS-7 is truncated as is conventional at degree 13 while the other models are visualized after truncating at degree 30. Our results show slightly higher power and more detailed structures, particularly the maximum of the marginal posterior. There are similarities to structures seen in other studies which have attempted to infer the core field above degree 13 \citep[e.g.][]{kalmag, AubertSepModel2020}. In particular the strong flux patch in the equatorial Atlantic is split into two as also seen by \cite{kalmag} while e.g. above and slightly to the west of this patch, small scale patches are present which were also seen in \cite{AubertSepModel2020}. The difference seen in the maximum of the marginal posterior map compared to the equivalent LSQ and posterior mean solutions indicate the presence of non-Gaussian features in the posterior realizations. 

Although the differences between the maps of the maximum of the marginal posterior and the posterior mean are minor, and less than the differences between either of them and traditional spherical harmonic-based models such as CHAOS-7, there are nevertheless some interesting features.  We note that the maximum of the marginal posterior shows higher amplitude features in the regions under the South Atlantic south-west of Africa; such features are important for understanding recent changes in the South Atlantic weak field anomaly at Earth's surface \citep{Finlay_2020}.  Higher amplitude features are seen in the central Pacific region, around latitude 15 degrees South, longitude 100 degrees West. Finally there is a noticeable East-West elongation of a positive flux feature South East of Madagascar around latitude 30 degrees South and 60 degrees East.  Overall the map of the maximum of the marginal posterior shows generally sharper features than the map of the posterior mean or that from CHAOS-7.

\begin{figure}[ht!]
	\centering
	\includegraphics[scale=.75]{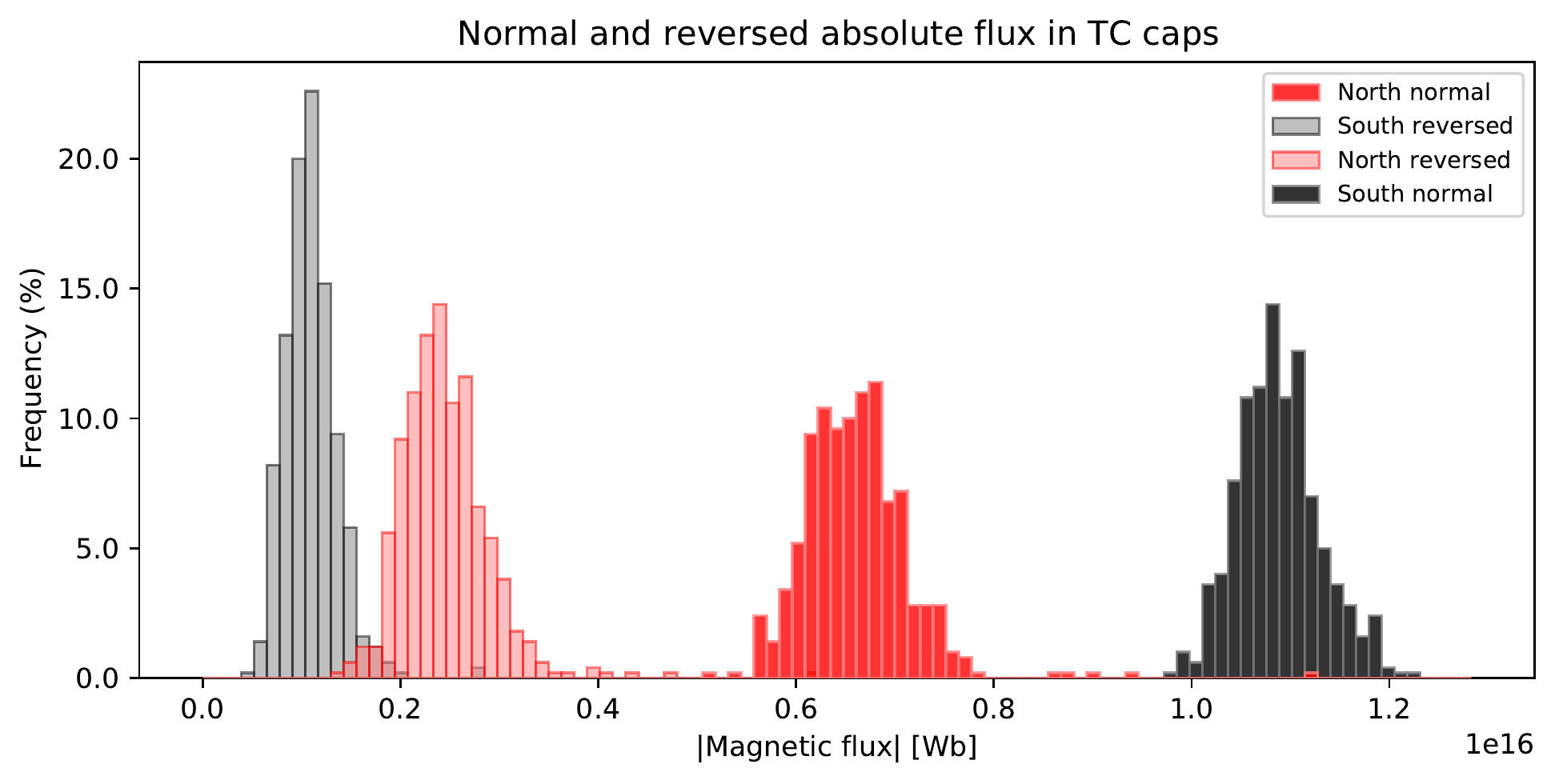}
    \caption{Distributions of absolute magnetic flux for normal and reversed values in the tangent cylinder caps at the core-mantle boundary. Each posterior realization has positive and negative contributions to the radial field within these regions, separately integrating the absolute values of positive and negative parts leads to the values reported here.}
    \label{fig:TC_dist}
\end{figure}

An example of a probabilistic investigation of CMB radial field structures is shown in Figure \ref{fig:TC_dist}.  This presents histograms of the integrated radial magnetic field inside the cylinder tangent to the Earth's solid inner core \citep{Livermore_2017}, separated into normal and reversed polarities, in the north and south hemispheres.  This analysis demonstrates the northern hemisphere has with high probability more reversed magnetic flux and weaker normal flux, a result of importance in geodynamo studies that was difficult to quantify with conventional field models.

\section{Discussion and conclusions}
\label{sec:discussion}
In the case studies presented we found that the posterior mean is close to the equivalent least-squares solution. Why then go to all the trouble of generating posterior realizations? A key point here is that marginal posterior distributions at particular locations can still be non-Gaussian (see e.g. Fig. \ref{fig:core_synth_local}) and hence are not necessarily well described by the posterior mean and variance. The importance of this has previously been highlighted in the Cartesian case \citep{visim} and will doubtless also prove crucial for some applications in spherical geometry, particularly when the prior model distributions are strongly non-Gaussian.

In the presented applications we have routinely transformed from the simulated grids in physical space to spherical harmonic representations.  This was found to be useful for comparisons with existing geomagnetic field models and for visualization, but care is needed with this procedure.  The transform from the grid in physical space to spherical harmonics is only exact for real square-integrable functions and when the spherical harmonic degree of the underlying function is limited to the level of chosen Gauss-Legendre quadrature grid \cite[e.g.][]{shtools}.  We observe a smoothing/loss of power on the grid scale compared to the originally simulated values in some of the results presented here.  This is acceptable if one wishes to compare models only up to some specific spherical harmonic degree, but for applications with covariance functions that allow discontinuities between neighbouring grid points, it is recommended to work instead with the simulated grids in physical space. There may be important advantages to working directly in the physical domain because the a priori covariance information can then be allowed to vary with position.  For geophysical problems involving Earth's lithosphere and upper mantle it may for instance be important to allow different covariance models for positions in the continents versus oceans, or to use locally defined information based on auxiliary variables such as geological composition or features. Allowing spatial variations in the a-priori covariance models is an obvious next step for the framework presented here.  The presented geomagnetic application was somewhat limited in the sense it involved only sources at one depth and ignored any time dependence. The extension to sources at multiple  depths (ideally with independently specified prior information) can be achieved by superposing the sources and visiting the model parameters at all depths during the sequential simulation. Similarly the model grids could be extended to a sequence of times in order to account for time-dependent source processes, provided the necessary time-dependent covariance matrices are specified \cite[see e.g.][]{Gillet_2013, ropp_sequential, kalmag}.

A major limitation of the present implementation of the SDSSIM algorithm is the use of two-point statistics (covariances) for describing the a-priori conditional relationship between the model parameters. Given the complexity of natural phenomena on the sphere
, these are not capable of fully capturing all the essential details.  In order to move beyond this limitation, similar algorithms in Cartesian geometry have utilized multiple-point statistics \citep{snesim, quick_sample}.  Use of multiple-point statistics in spherical geometry is not yet well developed, but would certainly be of interest for improving on the results obtained here. 

The SDSSIM scheme is in principle applicable to a wide variety of problems involving linear inversion or interpolation on a sphere. For example, possible applications could involve meteorological data such as the case presented by \cite{jun2008} where non-stationary covariance models are used to analyse global ozone levels or \cite{jeong2017} where isotropic and non-stationary covariance models are used with global surface temperature data. Extensions to 3D using grids at different radii and radial covariance functions is also possible, e.g. for inversion problems in seismology \citep{Meschede_2015} or gravity \citep{Save_etal_2016}. The success in such applications will rely on the availability of suitable prior information, for example in the form of training images or covariance fuctions. In such cases SDSSIM may allow for an improved exploitation of prior information and probabilistic descriptions of models in spherical geometry.

\section*{Funding sources}
This study was funded by the European Research Council (ERC) under the European Union’s Horizon 2020 research and innovation programme (Grant agreement No. 772561).

\section*{Computer Code Availability}
\begin{itemize}
    \item Name of code: spherical\_direct\_sequential\_simulation
    \item Developers: Mikkel Otzen
    \item Contact details: DTU Space, Centrifugevej 356, 2800 Kgs. Lyngby, Denmark, +4527576416, mikotz@space.dtu.dk
    \item Year first available: 2021
    \item Hardware required: Code was tested on a modern PC with 32GB RAM
    \item Software required: Python 3.6+ with the numpy, pyshtools, scikit-learn, matplotlib, and scipy packages
    \item Program language: Python
    \item Program size: $332$MB
    \item Details on how to access the source code: Available at \url{github.com/mikkelotzen/spherical_direct_sequential_simulation}
\end{itemize}

\bibliographystyle{cas-model2-names}
\bibliography{seqsim}

\end{document}